\documentclass[journal,12pt,onecolumn]{IEEEtran}
\usepackage{setspace}
\IEEEoverridecommandlockouts

\usepackage[colorlinks,urlcolor=blue,linkcolor=blue,citecolor=blue]{hyperref}
\usepackage{color,array,amsthm}
\usepackage{graphicx}
\usepackage{amsfonts}
\usepackage{amssymb}
\usepackage{amsmath}
\usepackage{indentfirst} 
\usepackage{booktabs} 
\usepackage{subfigure}
\usepackage{float}
\usepackage{CJK}
\usepackage{cite}
\usepackage[numbers,sort&compress]{natbib}
\usepackage{natbib}

\begin{document}

\title{Space-Time Adaptive Processing Using Random Matrix Theory Under Limited Training Samples} 

\author{Di~Song,
	Shengyao~Chen,
	Feng~Xi,
	and Zhong~Liu
%	\thanks{This work was supported in part by the National Natural Science Foundation of China under Grants 6217011933. (Corresponding author: S. Chen)}
%	\thanks{D. Song, S. Chen, F. Xi and Z. Liu are with the School of Electronic and Optical Engineering, Nanjing University of Science and Technology, Nanjing, China. (e-mail: 1742740029@qq.com; chenshengyao@njust.edu.cn; xifeng@njust.edu.cn; eezliu@njust.edu.cn)}
}

%%\editor{Mentions of supplemental materials and animal/human rights statements can be included here.}
%%\supplementary{Color versions of one or more of the figures in this article are available online at {http://ieeexplore.ieee.org}.}

\markboth{SONG ET AL.}{SPACE-TIME ADAPTIVE PROCESSING USING RANDOM MATRIX THEORY}
\maketitle

\begin{abstract}Space-time adaptive processing (STAP) is one of the most effective approaches to suppressing ground clutters in airborne radar systems. It basically takes two forms, i.e., full-dimension STAP (FD-STAP) and reduced-dimension STAP (RD-STAP). When the numbers of clutter training samples are less than two times their respective system degrees-of-freedom (DOF), the performances of both FD-STAP and RD-STAP degrade severely due to inaccurate clutter estimation. To enhance STAP performance under the limited training samples, this paper develops a STAP theory with random matrix theory (RMT). By minimizing the output clutter-plus-noise power, the estimate of the inversion of clutter plus noise covariance matrix (CNCM) can be obtained through optimally manipulating its eigenvalues, and thus producing the optimal STAP weight vector. Two STAP algorithms, FD-STAP using RMT (RMT-FD-STAP) and RD-STAP using RMT (RMT-RD-STAP), are proposed. It is found that both RMT-FD-STAP and RMT-RD-STAP greatly outperform other-related STAP algorithms when the numbers of training samples are larger than their respective clutter DOFs, which are much less than the corresponding system DOFs. Theoretical analyses and simulation demonstrate the effectiveness and the performance advantages of the proposed STAP algorithms.
\end{abstract}


\section{INTRODUCTION}
S{\scshape pace}-time adaptive processing (STAP) was first proposed  almost 50 years ago \cite{r1}, and since then it has been actively investigated by the radar community due to its strong capability in suppressing clutter \cite{r2,r3,r4}. Traditional STAP algorithms are implemented in either full dimension \cite{r3,r4,r5,r6} or reduced dimension mode \cite{r7,r8,r9,r10,r11,r12,r13,r14}. STAP in full dimension or full-dimension STAP (FD-STAP) can achieve the maximal signal-to-clutter plus noise ratio (SCNR) improvement and provide reliable detectability of moving targets. To do so, FD-STAP should fully take into account of the system degrees-of-freedom (DOF), which is defined as the product of the number of antenna elements and the number of pulses per coherent processing interval (CPI) \cite{r3}, \cite{r5} in the system design. In addition, there must be sufficient independent identical distributed (IID) clutter training samples to estimate the clutter plus noise covariance matrix (CNCM) of the cell under test (CUT) \cite{r3,r4,r5,r6}. In general, to gain the maximal SCNR improvement, one needs the number of the training samples much larger than the system DOFs.  To limit the SCNR loss to 3dB, we need to have twice as many IID clutter samples as that of the system DOFs, i.e., the RMB rule \cite{r15}. However, practically, the obtainable IID clutter training samples could be much less than that of the system DOFs \cite{r3,r4,r5}. Therefore, the RMB rule cannot be satisfied and the FD-STAP performance would get significantly deteriorated. As a result, STAP in reduced dimension or reduced-dimension STAP (RD-STAP) \cite{r7,r8,r9,r10,r11,r12,r13,r14} was developed with much reduced training sample requirement. RD-STAP tries to satisfy the RMB rule with the reduced system DOFs and thus has better performance under insufficient training data. In addition, the computational cost of RD-STAP is much lower than that of FD-STAP \cite{r4}.

RD-STAP algorithms, including the extended factored approach (EFA) \cite{r7}, \cite{r14}, the auxiliary channel processing \cite{r13} and the joint domain localized \cite{r8}, \cite{r11}, implement the reduction of the system DOFs through linear transformation on training samples. In a radar system, the transformation is often designed to be fixed in a predesigned clutter scenario. When the scenarios deviate from the predesigned ones, the RMB rule may not be satisfied and the RD-STAP performance becomes poor like a FD-STAP \cite{r3,r4}.

It should be noted that no matter what dimension STAP works in, it is still needed for the system to estimate the CNCM from the training samples and calculate its inversion \cite{r3}. A general scheme is to perform the maximum likelihood estimation with the training samples \cite{r4}. If with a small number of training samples the RMB rule cannot be satisfied, the estimated CNCM and its inversion deviate from the ideal ones and results in poor STAP performance \cite{r15}. In this paper, we take a different strategy for the clutter estimation by manipulating the eigenvalues of the CNCM from the training samples using the random matrix theory (RMT).

RMT originates from the quantum mechanics \cite{r16} and is mainly used to study the asymptotic behavior of the empirical spectral distribution of different random matrix models as their dimensions goes into infinity \cite{r17,r18,r19,r20,r21,r22,r23,r24}. It has been applied to estimate the eigenvalues and eigenvectors of the sample correlation/covariance matrices constructed from the finite observations \cite{r20}, \cite{r25,r26,r27,r28,r29,r30,r31,r32,r33,r34,r35,r36}. It is found that the estimation is consistent, not only when the sample size increases without bound for a fixed observation dimension, but also when the observation dimension increases to infinity at the same rate as the sample size increases \cite{r33,r34,r35}. This finding has found many applications in signal processing for communication, radar, sonar and so on \cite{r36,r37,r38,r39,r40,r41,r42,r43,r44,r45,r46}. Recently, \cite{r43} applies the theory to directly estimate the inverse covariance matrix for spatial beamforming and shows superior performance under high dimension and finite training samples. Motivated by this work, this paper is set to enhance the STAP performance using RMT. 

The CNCM eigenvalues consists of two parts, one corresponding to the noise and the other to the clutter \cite{r3}. The noise-related eigenvalues consist of the large portion of the CNCM eigenvalues with their values equal to the noise power. The clutter-related eigenvalues make up a small portion of the eigenvalues with their values much larger than the noise-related ones. The eigenvalue distribution is much like that of spiked covariance model defined as a low-rank perturbation of an identity matrix in RMT \cite{r47}. It is shown in \cite{r47,r48,r49,r50,r51} that for such a model, a consistent estimation on its eigenvalues can be obtained with small-size samples. This property has been exploited to improve the performance of spatial beamforming \cite{r43} and spatial spectrum estimation \cite{r50}. Similarly, the theory can be used to improve the estimation accuracy of the CNCM clutter-related eigenvalues with small training samples, which is essential to enhance STAP performance. References \cite{r52} and \cite{r53} studied the application of the model to FD-STAP with direct estimation of CNCM eigenvalues.

This paper studies STAP enhancement using RMT with limited training samples \cite{r3,r4,r5,r6}. Different from \cite{r52} and \cite{r53}, the proposed STAP estimates the eigenvalues of inverse CNCM by the spiked covariance model. Two types of STAP algorithms, FD-STAP using RMT (RMT-FD-STAP) and RD-STAP using RMT (RMT-RD-STAP), are developed with RMT. For FD-STAP, the noise covariance matrix is an identity matrix and it is straightforward to estimate the clutter-related eigenvalues by using RMT-based technique in \cite{r43}, \cite{r45}. For RD-STAP, the noise covariance matrix is not an identity one and the technique in \cite{r43}, \cite{r45} is not applicable. However, it is found that the idea in FD-STAP is still efficient for RD-STAP. The clutter-related eigenvalues in RD-STAP can be estimated by assuming that the noise covariance matrix is an identity matrix. Although there are no theoretical guarantees, the simulations in Section IV have confirmed the assertion. In addition, we conduct extensive simulations to verify the effectiveness of the RMT-based STAP. It is found that RMT-FD-STAP and RMT-RD-STAP greatly outperform FD-STAP and RD-STAP, respectively.

This paper is organized as follows. Section II introduces the echo model for airborne radar and the fundamentals of STAP. Section III presents the RMT-FD-STAP and RMT-RD-STAP algorithms and their computational costs. Numerical simulations are performed in section IV. The conclusions of this work are drawn in section V.

${\textit{Notation}}$: ${{\textbf{I}}_{N}}$ is defined as the identity matrix. Boldface uppercase letters denote the matrices, and boldface lowercase letters denote the vectors. $\otimes $  represents the Kronecker product. ${{(\cdot )}^{T}}$ and ${{(\cdot )}^{H}}$ represent the transpose and Hermitian transpose, respectively. $\mathbb{C}$  represents the sets of complex values. $\mathbb{E}\mathrm{(}\cdot \mathrm{)}$  denotes expectation operation.  $\left\lfloor \cdot  \right\rfloor $ indicates rounding to the nearest integer.

\section{BACKGROUND}

\subsection{Echo Model of Airborne Radar and Optimal STAP}

Consider a side-looking airborne pulsed-Doppler radar system equipped with a uniform linear array (ULA) of $N$ array elements with the inter-element spacing $\Delta =\lambda/2$ and the wavelength $\lambda$. Let $H$ and $V$ denote the height and velocity of the platform, respectively, and $\psi $, $\theta$ and $\varphi$ denote the cone, azimuth and elevation angle, respectively. Assume that the radar transmits $K$ pulses at a constant pulse repetition frequency (PRF) ${{f}_{r}}$ during a CPI. Then under the assumption of no range ambiguity and no internal clutter motion, the received signal  $\textit{\textbf{x}}_{l}\in {\mathbb{C}}^{NK}$ of the $l$-th range cell or the CUT can be represented as \cite{r3}
\begin{equation}
\textit{\textbf{x}}_{l}=\textit{\textbf{s}}+\textit{\textbf{c}}_{l}+\textit{\textbf{n}},
\end{equation}
where $\textit{\textbf{s}}$ denotes target echo,  $\textit{\textbf{n}}\sim \mathcal{C}\mathcal{N}\left( \textbf{0},\sigma _{n}^{2}{\textbf{I}_{NK}} \right)$ denotes the complex white Gaussian noise with the noise power $\sigma _{n}^{2}$, and  $\textit{\textbf{c}}_{l}$ denotes the clutter echo,
\begin{equation}
\textit{\textbf{c}}_{l}=\sum\limits_{i=1}^{{{N}_{c}}}{{{\Gamma }_{i}}{\textit{\textbf{a}}}\left( f_{ci}^{t},f_{ci}^{s} \right)}.
\end{equation}
In (2),  ${{N}_{c}}$ is the number of clutter patches evenly divided in azimuth,  ${{\Gamma }_{i}}$ denotes the complex amplitude of the $i$-th clutter patch, and $\textit{\textbf{a}}\left( f_{ci}^{t},f_{ci}^{s} \right)={{\textit{\textbf{a}}}_{t}}\left( f_{ci}^{t} \right)\otimes {{\textit{\textbf{a}}}_{s}}\left( f_{ci}^{s} \right)\in {{\mathbb{C}}^{NK}}$  denotes normalized spatial-temporal steering vector of the $i$-th clutter patch and ${{\textit{\textbf{a}}}_{t}}\left( f_{ci}^{t} \right)\in {{\mathbb{C}}^{K}}$  and  ${{\textit{\textbf{a}}}_{s}}\left( f_{ci}^{s} \right)\in {{\mathbb{C}}^{N}}$ denote the temporal and spatial steering vectors, respectively, 
\begin{equation}
\begin{aligned}
  & {{\textit{\textbf{a}}}_{t}}\left( f_{ci}^{t} \right)={{\left[ 1,{{e}^{j2\pi f_{ci}^{t}}},\cdots ,{{e}^{j2\pi f_{ci}^{t}(K-1)}} \right]}^{T}} \\ 
& {{\textit{\textbf{a}}}_{s}}\left( f_{ci}^{s} \right)={{\left[ 1,{{e}^{j2\pi f_{ci}^{s}}},\cdots ,{{e}^{j2\pi f_{ci}^{s}(N-1)}} \right]}^{T}}. \\
\end{aligned}
\end{equation}
In (3),  $f_{ci}^{t}$ and $f_{ci}^{s}$  denote normalized temporal and spatial frequency, respectively,
\begin{equation}
\begin{aligned}
  & f_{ci}^{t}=\frac{2V}{\lambda {{f}_{r}}}\cos {{\psi }_{i}}=\frac{2V}{\lambda {{f}_{r}}}\cos {{\theta }_{i}}\cos {{\varphi }_{l}} \\ 
& f_{ci}^{s}=\frac{\Delta }{\lambda }\cos {{\psi }_{i}}=\frac{\Delta }{\lambda }\cos {{\theta }_{i}}\cos {{\varphi }_{l}}.\\ 
\end{aligned}
\end{equation}
Then the CNCM $\textbf{\textit{{R}}}$ of the  $l$-th range cell can be expressed from (1) as
\begin{equation}
\begin{aligned}
\textbf{\textit{{R}}}
&=\mathbb{E}\left( {{\textbf{\textit{{x}}}}_{l}}\textbf{\textit{{x}}}_{l}^{H} \right)\\
&=\sum\limits_{i=1}^{{{N}_{c}}}{{{\left| {{\Gamma }_{i}} \right|}^{2}}{\textbf{\textit{{a}}}}\left( f_{ci}^{t},f_{ci}^{s} \right){{\textbf{\textit{{a}}}}^{H}}\left( f_{ci}^{t},f_{ci}^{s} \right)}+\sigma _{n}^{2}{{\textbf{{I}}}_{NK}}.
\end{aligned}
\end{equation}

Under the linear constraint minimum variance (LCMV) criterion \cite{r5}, the optimal STAP is defined as
\begin{equation}
\left\{ \begin{aligned}
& \min  P\left( \textbf{\textit{w}} \right)={{\textbf{\textit{w}}}^{H}}{\textbf{\textit{R}}}{\textbf{\textit{w}}} \\ 
& s.t. \quad {{\textbf{\textit{w}}}^{H}}{\textbf{\textit{a}}}\left( f_{0}^{t},f_{0}^{s} \right)=1, \\ 
\end{aligned} \right.
\end{equation}
where  $P\left( \textbf{\textit{{w}}} \right)$ denotes the output clutter-plus-noise power, $ {\textbf{\textit{{a}}}}\left( f_{0}^{t},f_{0}^{s} \right)\in {{\mathbb{C}}^{NK}}$ denotes the spatial-temporal steering vector of the target with   $f_{0}^{t}$ and $f_{0}^{s}$  as the normalized temporal and spatial frequency. The optimal STAP weight vector ${\textbf{\textit{{w}}}\in {{\mathbb{C}}^{NK}}}$  can be derived as \cite{r5}
\begin{equation}
{{\textbf{\textit{{w}}}}^{opt}}=\frac{{{\textbf{\textit{{R}}}}^{-1}}{\textbf{\textit{{a}}}}\left( f_{0}^{t},f_{0}^{s} \right)}{{{\textbf{\textit{{a}}}}^{H}}\left( f_{0}^{t},f_{0}^{s} \right){{\textbf{\textit{{R}}}}^{-1}}{\textbf{\textit{{a}}}}\left( f_{0}^{t},f_{0}^{s} \right)},
\end{equation}
from which, the minimal clutter-plus-noise power is given by
\begin{equation}
P\left( {{\textbf{\textit{{w}}}}^{opt}} \right)=\frac{1}{{{\textbf{\textit{{a}}}}^{H}}\left( f_{0}^{t},f_{0}^{s} \right){{\textbf{\textit{{R}}}}^{-1}}{\textbf{\textit{{a}}}}\left( f_{0}^{t},f_{0}^{s} \right)}.
\end{equation}

\subsection{ Sample-Based STAP}

For the implementation of (7), it is fundamental to know the CNCM ${\textbf{\textit{{R}}}}$  and its inversion. However, it is not practical to obtain the matrix ${\textbf{\textit{{R}}}}$. In practice, the matrix ${\textbf{\textit{{R}}}}$   is estimated from finite training samples \cite{r3,r4,r5},
\begin{equation}
{{\textbf{\textit{{R}}}}_{L}}=\sum\limits_{l=1}^{L}{{{\textbf{\textit{{x}}}}_{l}}{\textbf{\textit{{x}}}}_{l}^{\mathrm{H}}},
\end{equation}
where ${{\textbf{\textit{{R}}}}_{L}}$  is called sample CNCM with $L$ training samples. In such case, the optimal weight vector (7) is approximated as 
\begin{equation}
{\textbf{\textit{{w}}}}_{L}^{opt}=\frac{{\textbf{\textit{{R}}}}_{L}^{-1}{\textbf{\textit{{a}}}}\left( f_{0}^{t},f_{0}^{s} \right)}{{{\textbf{\textit{{a}}}}^{H}}\left( f_{0}^{t},f_{0}^{s} \right){\textbf{\textit{{R}}}}_{L}^{-1}{\textbf{\textit{{a}}}}\left( f_{0}^{t},f_{0}^{s} \right)},
\end{equation}
where ${\textbf{\textit{{w}}}}_{L}^{opt}$  is called the adaptive weight vector of the FD-STAP. The output clutter-plus-noise power by the weight vector (10) is given by
\begin{equation}
P\left( {\textbf{\textit{{w}}}}_{L}^{opt} \right)=\frac{{{{\textbf{\textit{{a}}}}}^{H}}\left( f_{0}^{t},f_{0}^{s} \right){\textbf{\textit{{R}}}}_{L}^{-1}{\textbf{\textit{{R}}}}{\textbf{\textit{{R}}}}_{L}^{-1}{\textbf{\textit{{a}}}}\left( f_{0}^{t},f_{0}^{s} \right)}{{{\left[ {{\textbf{\textit{{a}}}}^{H}}\left( f_{0}^{t},f_{0}^{s} \right){\textbf{\textit{{R}}}}_{L}^{-1}{\textbf{\textit{{a}}}}\left( f_{0}^{t},f_{0}^{s} \right) \right]}^{2}}}.
\end{equation}

It is seen from (11) that the FD-STAP will have good performance if ${\textbf{\textit{{R}}}}_{L}^{-1}$  well approximates ${\textbf{\textit{{R}}}}^{-1}$. However, ${\textbf{\textit{{R}}}}_{L}^{-1}$  is estimated from finite training samples and deviates from ${\textbf{\textit{{R}}}}^{-1}$  as the number of training samples decreases. This deviation will result in higher output clutter-plus-noise power by (11) than theoretical one by (8), especially when the number of training samples is not larger than the system DOFs, $L\le NK$. Define a constant $c\in (0,1)$. Then from \cite{r37,r38,r39}, it is known that ${P\left( {\textbf{\textit{{w}}}}_{L}^{opt} \right)}/{P\left( {{\textbf{\textit{{w}}}}^{opt}} \right)}\to {1}/{\left( 1-c \right)}$  as  $NK,L\to \infty $ with ${{c}_{N}}={NK}/{L}\to c$, which coincides with the RMB rule, i.e., the FD-STAP performance by (10) has less than 3dB loss when the number of the training samples is greater than two times system DOFs, $L\ge 2NK$. However, in real implementation, it is a common case that $L<NK$  \cite{r3,r4,r5,r6}. The FD-STAP by (10) will have much poorer performance than that by (7).

To reduce the requirement of (10) on the number of training samples, RD-STAP \cite{r5,r6,r7} is developed, which produces a secondary data by projecting the $NK$-dimensional signal vector $\textit{\textbf{x}}_{l}$  into a $M$-dimensional one via linear transformation matrix ${\textbf{\textit{{T}}}}\in {{\mathbb{C}}^{NK\times M}}$,
\begin{equation}
{{\tilde{\textit{\textbf{x}}}}_{l}}={{\textit{\textbf{T}}}^{H}}{{\textit{\textbf{x}}}_{l}}\in {{\mathbb{C}}^{M}},
\end{equation}
where $M$  denotes the reduced-dimensional system DOFs and is much smaller than the system DOFs, $M\ll NK$. Then it can be predicted that the STAP with the data (12) will have good performance when $L\ge 2M$  in terms of the RMB rule. 

The CNCM and the target spatial-temporal steering vector in the reduced-dimensional data are given as
\begin{equation}
\begin{aligned}
& {{{\textit{\textbf{R}}}}_{rd}}=\mathbb{E}\left( {{{\tilde{{\textit{\textbf{x}}}}}}_{l}}\tilde{{\textit{\textbf{x}}}}_{l}^{H} \right)={{{\textit{\textbf{T}}}}^{H}}{{\textit{\textbf{R}}}}{{\textit{\textbf{T}}}}\in {{\mathbb{C}}^{M\times M}} \\
& {{{{{\textit{\textbf{a}}}}}}_{rd}}\left( f_{0}^{t},f_{0}^{s} \right)={{{\textit{\textbf{T}}}}^{H}}{{\textit{\textbf{a}}}}\left( f_{0}^{t},f_{0}^{s} \right)\in {{\mathbb{C}}^{M}} .\\ 
\end{aligned}
\end{equation}
Under the LCMV criterion, the RD-STAP is defined as 
\begin{equation}
\left\{ \begin{aligned}
& \min {{P}_{rd}}\left( {\textit{\textbf{w}}}_{rd} \right)={\textit{\textbf{w}}}_{rd}^{H}{{\textit{\textbf{R}}}_{rd}}{{\textit{\textbf{w}}}_{rd}} \\ 
& s.t.     {\textit{\textbf{w}}}_{rd}^{H}{{\textit{\textbf{a}}}_{rd}}\left( f_{0}^{t},f_{0}^{s} \right)=1. 
\end{aligned} \right.
\end{equation}
where ${\textit{\textbf{w}}}_{rd}\in {{\mathbb{C}}^{M}}$  denotes the RD-STAP weight vector. The optimal RD-STAP weight vector are given as
\begin{equation}
{\textit{\textbf{w}}}_{rd}^{opt}=\frac{{\textit{\textbf{R}}}_{rd}^{-1}{{\textit{\textbf{a}}}_{rd}}\left( f_{0}^{t},f_{0}^{s} \right)}{{\textit{\textbf{a}}}_{rd}^{H}\left( f_{0}^{t},f_{0}^{s} \right){\textit{\textbf{R}}}_{rd}^{-1}{{\textit{\textbf{a}}}_{rd}}\left( f_{0}^{t},f_{0}^{s} \right)}.
\end{equation}
When the ${{\textit{\textbf{R}}}_{rd}}$  is obtained from the finite samples, the adaptive weight vector of the RD-STAP is
\begin{equation}
{\textit{\textbf{w}}}_{Lrd}^{opt}=\frac{{\textit{\textbf{R}}}_{Lrd}^{-1}{{\textit{\textbf{a}}}_{rd}}\left( f_{0}^{t},f_{0}^{s} \right)}{{\textit{\textbf{a}}}_{rd}^{H}\left( f_{0}^{t},f_{0}^{s} \right){{\textit{\textbf{R}}}}_{Lrd}^{-1}{{\textit{\textbf{a}}}_{rd}}\left( f_{0}^{t},f_{0}^{s} \right)}.
\end{equation}
where  ${{\textit{\textbf{R}}}_{Lrd}}={{\textit{\textbf{T}}}^{H}}{{\textit{\textbf{R}}}_{L}}{\textit{\textbf{T}}}\in {{\mathbb{C}}^{M\times M}}$ is the sample CNCM in reduced dimension. Similar to (11), the output clutter-plus-noise power by the weight vector (16) is given as
\begin{equation}
{{P}_{rd}}\left( {\textit{\textbf{w}}}_{Lrd}^{opt} \right)\text{=}\sigma _{n}^{2}\frac{{\textit{\textbf{a}}}_{rd}^{H}\left( f_{0}^{t},f_{0}^{s} \right){\textit{\textbf{R}}}_{Lrd}^{-1}{{\textit{\textbf{R}}}_{rd}}{\textit{\textbf{R}}}_{Lrd}^{-1}{{\textit{\textbf{a}}}_{rd}}\left( f_{0}^{t},f_{0}^{s} \right)}{{{\left[ {\textit{\textbf{a}}}_{rd}^{H}\left( f_{0}^{t},f_{0}^{s} \right){\textit{\textbf{R}}}_{Lrd}^{-1}{{\textit{\textbf{a}}}_{rd}}\left( f_{0}^{t},f_{0}^{s} \right) \right]}^{2}}}.
\end{equation}

\subsection{ Motivation by Introducing RMT in STAP}
In this paper, we study the consistent estimation of ${{\textit{\textbf{R}}}^{-1}}$  and ${\textit{\textbf{R}}}_{rd}^{-1}$  from the finite training samples, especially in the case of $L<NK$. 

For the inverse matrix ${{\textit{\textbf{R}}}^{-1}}$, let us consider the eigen-decomposition of the matrix ${\textit{\textbf{R}}}$ 
\begin{equation}
\begin{aligned}
 {\textit{\textbf{R}}}
&=\sum\limits_{i=1}^{Q}{\sigma _{n}^{2}\left( {{\rho }_{i}}+1 \right){{{\textbf{v}}}_{i}}{{\textbf{v}}}_{i}^{H}}+\sum\limits_{i=Q+1}^{NK}{\sigma _{n}^{2}{{{\textbf{v}}}_{i}}{{\textbf{v}}}_{i}^{H}} \\ 
&=\sigma _{n}^{2}\left( \sum\limits_{i=1}^{Q}{{{\rho }_{i}}{{{\textbf{v}}}_{i}}{{\textbf{v}}}_{i}^{H}}+{{{\textbf{I}}}_{NK}} \right) \\ 
\end{aligned}
\end{equation}
where  ${{\textbf{v}}_{i}}$ is the $i$-th eigenvector with the corresponding eigenvalue  $\sigma _{n}^{2}\left( {{\rho }_{i}}+1 \right)$ and ${{\rho }_{1}}\ge \cdots \ge {{\rho }_{Q}}\ge \sigma _{n}^{2}$, and  $Q=\left\lfloor N+\left( K-1 \right)\xi  \right\rfloor$ is the clutter rank or clutter DOFs with  $\xi ={2V}/{\left( \Delta {{f}_{r}} \right)}$ as the slope of the clutter ridge \cite{r3}. The clutter rank  $Q$ varies as the slop parameter $\xi$  changes. The STAP performance becomes poor as the clutter rank $Q$  increases \cite{r3,r4,r5,r6}.

Notice that $Q$  is far less than the system DOFs $NK$  in practice \cite{r3}. Then the matrix ${\textit{\textbf{R}}}$  (18) is a low-rank perturbation of the identity matrix and obeys the so-called spiked covariance model in RMT. In the area of RMT, the model has acquired extensive study on its statistical behaviors \cite{r47,r48,r49} and the consistent estimation theory has been established on estimating the isolated eigenvalues of the spiked covariance model from finite training samples \cite{r47,r48,r49,r50,r51}. In particular, the spike covariance matrix estimation can be improved by managing the sample eigenvalues while keeping the sample eigenvectors \cite{r27,r28,r29,r30}, \cite{r42}, \cite{r43}, \cite{r52}. Then this theory can be leveraged to improve the estimation of ${{\textit{\textbf{R}}}^{-1}}$  and in turn enhance the STAP performance.

This theory is also suitable for the estimation of ${\textit{\textbf{R}}}_{rd}^{-1}$.

\section{STAP using RMT}

\subsection{ Full-dimension STAP Using RMT}
This subsection develops a full-dimension STAP using RMT. Firstly, the RMT-FD-STAP problem with the spiked covariance model is defined; then its asymptotic deterministic equivalence is given and the optimal solution is derived; and finally an estimated adaptive weight vector is optimized.

\subsubsection{FD-STAP Problem under the Spiked Covariance Model}
\
\newline
\indent 
Similar to (18), the matrix ${{\textit{\textbf{R}}}_{L}}$  can be decomposed as
\begin{equation}
{{\textit{\textbf{R}}}_{L}}=\sigma _{n}^{2}\sum\limits_{i=1}^{NK}{{{\mu }_{i}}{{\textbf{u}}_{i}}{\textbf{u}}_{i}^{H}},
\end{equation}
where  ${\textbf{u}}_{i}$ denotes the  ${i}$-th eigenvector of the matrix ${\textit{\textbf{R}}}_{L}$  with the corresponding eigenvalue  $\sigma _{n}^{2}{{\mu }_{i}}$ and ${{\mu }_{1}}\ge \cdots \ge {{\mu }_{NK}}$. Our task is to estimate  ${\textit{\textbf{R}}}^{-1}$ from the available ${\textit{\textbf{R}}}_{L}$. Then with (19), we can reconstruct the inverse CNCM by taking
\begin{equation}
{{\bar{\textit{\textbf{R}}}}^{-1}}=\frac{1}{\sigma _{n}^{2}}\sum\limits_{i=1}^{NK}{{{\eta }_{i}}{{\textbf{u}}_{i}}{\textbf{u}}_{i}^{H}},
\end{equation}
where  ${1}/{\sigma _{n}^{2}}{{\eta }_{i}}$ are the eigenvalues of  ${{\bar{\textit{\textbf{R}}}}^{-1}}$ and the parameters  ${{\eta }_{i}}$ are to be reconstructed.  From (18), the noise eigenvalues are all equal to the noise power. Then it is natural to set ${{\eta }_{Q+1}}=\cdots ={{\eta }_{NK}}=1$  and the inverse CNCM can be re-expressed as
\begin{equation}
{{\bar{\textit{\textbf{R}}}}^{-1}}=\frac{1}{\sigma _{n}^{2}}\left[ \sum\limits_{i=1}^{Q}{\left( {{\eta }_{i}}-1 \right){{\textbf{u}}_{i}}{\textbf{u}}_{i}^{H}}+{{\textbf{I}}_{NK}} \right].
\end{equation}

For simplicity, define ${{h}_{i}}={{\eta }_{i}}-1$. The (21) is simplified as
\begin{equation}
{{\bar{\textit{\textbf{R}}}}^{-1}}\left( \textbf{h} \right)=\frac{1}{\sigma _{n}^{2}}\left( \sum\limits_{i=1}^{Q}{{{h}_{i}}{{\textbf{u}}_{i}}{\textbf{u}}_{i}^{H}}+{\textbf{I}}_{NK} \right),
\end{equation}
in which the inverse matrix  ${{\bar{\textit{\textbf{R}}}}^{-1}}$ is implicitly expressed as a function of the vector ${\textbf{h}}={{\left[ {{h}_{1}},\cdots ,{{h}_{Q}} \right]}^{T}}$ . The adaptive weight vector ${{\bar{\textit{\textbf{w}}}}^{opt}}$  by the inverse CNCM (22) is given as
\begin{equation}
{{\bar{\textit{\textbf{w}}}}^{opt}}\left( \bf{h} \right)=\frac{{{{\bar{\textit{\textbf{R}}}}}^{-1}}\left( {\textbf{h}} \right){\textit{\textbf{a}}}\left( f_{0}^{t},f_{0}^{s} \right)}{{{\textit{\textbf{a}}}^{H}}\left( f_{0}^{t},f_{0}^{s} \right){{{\bar{\textit{\textbf{R}}}}}^{-1}}\left( {\textbf{h}} \right){\textit{\textbf{a}}}\left( f_{0}^{t},f_{0}^{s} \right)}.
\end{equation}
Then the RMT-FD-STAP problem under the spiked covariance model can be defined as
\begin{equation}
{{\textbf{h}}^{*}}=\underset{\textbf{h}}{\mathop{argmin}}\,P\left( {{{\bar{\textit{\textbf{w}}}}}^{opt}}\left( {\textbf{h}} \right) \right),
\end{equation}
where $P\left( {{{\bar{\textit{\textbf{w}}}}}^{opt}}\left( {\textbf{h}} \right) \right)$ is given by %(25) (see the top line of the next page).
%\newcounter{mytempeqncnt}
%\begin{figure*}[!t]	
%	% ensure that we have normalsize text	
%	\normalsize	
%	% Store the current equation number.	
%	\setcounter{mytempeqncnt}{24}	
%	% Set the equation number to one less than the one	
%	% desired for the first equation here.	
%	% The value here will have to changed if equations	
%	% are added or removed prior to the place these	
%	% equations are referenced in the main text.	
%%	\setcounter{equation}{24}
	\begin{equation}
	\begin{aligned}
	P\left( {{{\bar{\textit{\textbf{w}}}}}^{opt}}\left( {\textbf{h}} \right) \right)
	&=\frac{{{\textit{\textbf{a}}}^{H}}\left( f_{0}^{t},f_{0}^{s} \right){{{\bar{\textit{\textbf{R}}}}}^{-1}}\left( {\textbf{h}} \right){\textit{\textbf{R}}}{{{\bar{\textit{\textbf{R}}}}}^{-1}}\left( {\textbf{h}} \right){\textit{\textbf{a}}}\left( f_{0}^{t},f_{0}^{s} \right)}{{{\left[ {{\textit{\textbf{a}}}^{H}}\left( f_{0}^{t},f_{0}^{s} \right){{{\bar{\textit{\textbf{R}}}}}^{-1}}\left( {\textbf{h}} \right){\textit{\textbf{a}}}\left( f_{0}^{t},f_{0}^{s} \right) \right]}^{2}}} \\ 
	& =\sigma _{n}^{2}\frac{{{\textit{\textbf{a}}}^{H}}\left( f_{0}^{t},f_{0}^{s} \right)\left( {{\textbf{I}}_{NK}}+\sum\limits_{q=1}^{Q}{{{h}_{q}}{{\textbf{u}}_{q}}{\textbf{u}}_{q}^{H}} \right)\left( {{\textbf{I}}_{NK}}+\sum\limits_{j=1}^{Q}{{{\rho }_{j}}{{\textbf{v}}_{j}}{\textbf{v}}_{j}^{H}} \right)\left( {{\textbf{I}}_{NK}}+\sum\limits_{i=1}^{Q}{{{h}_{i}}{{\textbf{u}}_{i}}{\textbf{u}}_{i}^{H}} \right){\textit{\textbf{a}}}\left( f_{0}^{t},f_{0}^{s} \right)}{{{\left[ {{\textit{\textbf{a}}}^{H}}\left( f_{0}^{t},f_{0}^{s} \right)\left( {{\textbf{I}}_{NK}}+\sum\limits_{i=1}^{Q}{{{h}_{i}}{{\textbf{u}}_{i}}{\textbf{u}}_{i}^{H}} \right){\textit{\textbf{a}}}\left( f_{0}^{t},f_{0}^{s} \right) \right]}^{2}}}  \\ 
	\end{aligned}
	\end{equation}
%\setcounter{equation}{25}
%% IEEE uses as a separator
%\hrulefill
%% The spacer can be tweaked to stop underfull vboxes.
%\vspace*{4pt}
%\end{figure*}

The optimal RMT-FD-STAP becomes to finding the optimal  ${{h}_{i}}$ or optimal ${{\eta }_{i}}$.

\subsubsection{	Asymptotic Deterministic Equivalent $\tilde{P}\left( {{{\bar{\textit{\textbf{w}}}}}^{opt}}\left( {\textbf{h}} \right) \right)$  and the Optimal ${{\tilde{\textbf{h}}}^{*}}$  }
\
\newline
\indent 
Note that it is difficult to give a closed solution of (24). To address the issue, the asymptotic properties of (25) under $NK,L\to \infty $  are considered. To conduct such an analysis, three assumptions are made:

{\bf{A.1}} As $NK,L\to \infty $,  ${NK}/{L}={{c}_{N}}\to c$ for a certain $c>0$.

{\bf{A.2}} The number of  $Q$ is fixed, and smaller than $NK$.

{\bf{A.3}} ${{\rho }_{1}}\ge \cdots \ge {{\rho }_{Q}}>\sqrt{c}$ and ${{\mu }_{1}}\ge \cdots \ge {{\mu }_{Q}}>{{\left( 1+\sqrt{c} \right)}^{2}}$.

For A.1, both  $NK$ and $L$  are assumed to be reasonably large. When $c>1$, the number of the training samples is less than the system DOFs, i.e., $L<NK$, the case in which we consider in this paper. A.2 is often taken in the airborne phased-array radar system and $Q<NK$  \cite{r3,r4,r5,r6}. The assumption ${{\rho }_{1}}\ge \cdots \ge {{\rho }_{Q}}>\sqrt{c}$  in A.3 is easily satisfied in practice because the clutter power is far higher than the noise power \cite{r3,r4,r5,r6}. For ${{\mu }_{1}}\ge \cdots \ge {{\mu }_{Q}}>{{\left( 1+\sqrt{c} \right)}^{2}}$, the number $L$  of the training samples is not less than the clutter DOFs,  $L\ge Q$. Otherwise ${{\mu }_{L+1}}=\cdots ={{\mu }_{Q}}=0<{{\left( 1+\sqrt{c} \right)}^{2}}$, and ${{\rho }_{i}}$  cannot be uniquely determined from  ${{\mu }_{i}}$ \cite{r47}, leading to the impossibility to recover the matrix ${\textit{\textbf{R}}}$  \cite{r33}, \cite{r34}, \cite{r47}.

The three assumptions follow \cite{r33,r34,r35} which are fundamental to derive the asymptotic properties of (25), as discussed in \cite{r43}, \cite{r45}. Then it can be shown for the FD-STAP that
\begin{equation}
\begin{aligned}
& \left| {\textbf{v}}_{i}^{H}{{\textbf{u}}_{j}}{\textbf{u}}_{j}^{H}{{\textbf{v}}_{i}}-{{s}_{i}}{{\delta }_{ij}} \right|\xrightarrow{a.s.}0 \qquad \text{     }i,j=1,\ldots ,Q \\ 
& \left| {{\textit{\textbf{a}}}^{H}}\left( f_{0}^{t},f_{0}^{s} \right){{\textbf{u}}_{i}}{\textbf{u}}_{i}^{H}{\textit{\textbf{a}}}\left( f_{0}^{t},f_{0}^{s} \right)-{{s}_{i}}{{k}_{i}} \right|\xrightarrow{a.s.}0 \\ 
& \left| {{\textit{\textbf{a}}}^{H}}\left( f_{0}^{t},f_{0}^{s} \right){{\textbf{u}}_{i}}{\textbf{u}}_{i}^{H}{{\textbf{v}}_{i}}{\textbf{v}}_{i}^{H}{\textit{\textbf{a}}}\left( f_{0}^{t},f_{0}^{s} \right)-{{s}_{i}}{{k}_{i}} \right|\xrightarrow{a.s.}0 \\ 
& \left| {{\textit{\textbf{a}}}^{H}}\left( f_{0}^{t},f_{0}^{s} \right){{\textbf{u}}_{i}}{\textbf{u}}_{i}^{H}{{\textbf{v}}_{i}}{\textbf{v}}_{i}^{H}{{\textbf{u}}_{i}}{\textbf{u}}_{i}^{H}{\textit{\textbf{a}}}\left( f_{0}^{t},f_{0}^{s} \right)-s_{i}^{2}{{k}_{i}} \right|\xrightarrow{a.s.}0 ,\\ 
\end{aligned}
\end{equation}
where ${{\delta }_{ij}}$  is the Kronecker-delta function, ${{s}_{i}}={\left( 1-{{{c}_{N}}}/{\rho _{i}^{2}}\; \right)}/{\left( 1+{{{c}_{N}}}/{{{\rho }_{i}}}\; \right)}\;$, and ${{k}_{i}}={{\textit{\textbf{a}}}^{H}}\left( f_{0}^{t},f_{0}^{s} \right){{\textbf{v}}_{i}}{\textbf{v}}_{i}^{H}{\textit{\textbf{a}}}\left( f_{0}^{t},f_{0}^{s} \right)$. In terms of (26), $P\left( {{{\bar{\textit{\textbf{w}}}}}^{opt}}\left( {\textbf{h}} \right) \right)$  has its asymptotic deterministic equivalent  $\tilde{P}\left( {{{\bar{\textit{\textbf{w}}}}}^{opt}}\left( {\textbf{h}} \right) \right)$ as
\begin{equation}
\begin{aligned}
\tilde{P}\left( {{{\bar{\textit{\textbf{w}}}}}^{opt}}\left( {\textbf{h}} \right) \right)
&=\frac{\sigma _{n}^{2}{{\textit{\textbf{a}}}^{H}}\left( f_{0}^{t},f_{0}^{s} \right){\textit{\textbf{a}}}\left( f_{0}^{t},f_{0}^{s} \right)}{{{\left[ {{\textit{\textbf{a}}}^{H}}\left( f_{0}^{t},f_{0}^{s} \right){\textit{\textbf{a}}}\left( f_{0}^{t},f_{0}^{s} \right)+\sum\limits_{i=1}^{Q}{{{h}_{i}}{{s}_{i}}{{k}_{i}}} \right]}^{2}}}+ \\ 
& \frac{\sigma _{n}^{2}\left( 2\sum\limits_{i=1}^{Q}{{{h}_{i}}{{s}_{i}}{{k}_{i}}}+\sum\limits_{i=1}^{Q}{{{k}_{i}}{{\rho }_{i}}}+2\sum\limits_{i=1}^{Q}{{{h}_{i}}{{\rho }_{i}}{{s}_{i}}{{k}_{i}}} \right)}{{{\left[ {{\textit{\textbf{a}}}^{H}}\left( f_{0}^{t},f_{0}^{s} \right){\textit{\textbf{a}}}\left( f_{0}^{t},f_{0}^{s} \right)+\sum\limits_{i=1}^{Q}{{{h}_{i}}{{s}_{i}}{{k}_{i}}} \right]}^{2}}} \\ 
& +\text{                       }\frac{\sigma _{n}^{2}\left( \sum\limits_{i=1}^{Q}{h_{i}^{2}{{s}_{i}}{{k}_{i}}}+\sum\limits_{i=1}^{Q}{h_{i}^{2}{{\rho }_{i}}s_{i}^{2}{{k}_{i}}} \right)}{{{\left[ {{\textit{\textbf{a}}}^{H}}\left( f_{0}^{t},f_{0}^{s} \right){\textit{\textbf{a}}}\left( f_{0}^{t},f_{0}^{s} \right)+\sum\limits_{i=1}^{Q}{{{h}_{i}}{{s}_{i}}{{k}_{i}}} \right]}^{2}}}, \\
\end{aligned}
\end{equation}
and
\begin{equation}
\left| P\left( {{{\bar{\textit{\textbf{w}}}}}^{opt}}\left( {\textbf{h}} \right) \right)-\tilde{P}\left( {{{\bar{\textit{\textbf{w}}}}}^{opt}}\left( {\textbf{h}} \right) \right) \right|\xrightarrow{a.s.}0.
\end{equation}
By replacing $P\left( {{{\bar{\textit{\textbf{w}}}}}^{opt}}\left( {\textbf{h}} \right) \right)$  with $\tilde{P}\left( {{{\bar{\textit{\textbf{w}}}}}^{opt}}\left( {\textbf{h}} \right) \right)$  in (24), we have the optimal ${{\tilde{\textbf{h}}}^{*}}={{\left[ \tilde{h}_{1}^{*},\cdots ,\tilde{h}_{Q}^{*} \right]}^{T}}$ by minimizing (27) as
\begin{equation}
\begin{aligned}
\tilde{h}_{i}^{*}
&=\frac{{{\rho }_{i}}+{{c}_{N}}}{\rho _{i}^{2}+{{\rho }_{i}}} 
%\centerdot \\
&\left[ \frac{\sum\limits_{q=1}^{Q}{\frac{{{c}_{N}}{{k}_{q}}}{{{\rho }_{q}}}}}{{{a}^{H}}\left( f_{0}^{t},f_{0}^{s} \right)a\left( f_{0}^{t},f_{0}^{s} \right)-\sum\limits_{q=1}^{Q}{{{k}_{q}}}+\sum\limits_{q=1}^{Q}{\frac{{{c}_{N}}{{k}_{q}}}{\rho _{q}^{2}}}}-{{\rho }_{i}} \right].
\end{aligned}
\end{equation}

\subsubsection{Estimating Optimal Adaptive Weight Vector ${{\hat{\textit{\textbf{w}}}}^{opt}}$  by Estimating  $\hat{h}_{i}^{*}$ }
\
\newline
\indent 
It is seen from (29) that the computation of the optimal values $\tilde{h}_{i}^{*}$  involve the unknown quantities ${\rho }_{i}$  and $k_{i}$. Note that under the assumption A.3, there are one-to-one maps between  ${{\mu }_{i}}$ and ${\rho }_{i}$, $i=1,\ldots ,Q$. With the maps, the unknown quantities ${\rho }_{i}$  and $k_{i}$  satisfy \cite{r47,r48,r49},
\begin{equation}
\begin{aligned}
& \left| {{\mu }_{i}}-1-{{\rho }_{i}}-\frac{{{c}_{N}}\left( 1+{{\rho }_{i}} \right)}{{{\rho }_{i}}} \right|\xrightarrow{a.s.}0 \\ 
& \left| {{k}_{i}}-\frac{1+{{{c}_{N}}}/{{{{\hat{\rho }}}_{i}}}\;}{1-{{{c}_{N}}}/{\hat{\rho }_{i}^{2}}\;}{{\textit{\textbf{a}}}^{H}}\left( f_{0}^{t},f_{0}^{s} \right){{\textbf{u}}_{i}}{\textbf{u}}_{i}^{H}{\textit{\textbf{a}}}\left( f_{0}^{t},f_{0}^{s} \right) \right|\xrightarrow{a.s.}0. \\ 
\end{aligned}
\end{equation}
Then the consistent estimate of the unknown quantities  ${\rho }_{i}$  and $k_{i}$  are given by
\begin{equation}
\begin{aligned}
& {{{\hat{\rho }}}_{i}}=\frac{{{\mu }_{i}}-1-{{c}_{N}}+\sqrt{{{\left( {{\mu }_{i}}-1-{{c}_{N}} \right)}^{2}}-4{{c}_{N}}}}{2} \\ 
& {{{\hat{k}}}_{i}}=\frac{1+{{{c}_{N}}}/{{{{\hat{\rho }}}_{i}}}\;}{1-{{{c}_{N}}}/{\hat{\rho }_{i}^{2}}\;}{{\textit{\textbf{a}}}^{H}}\left( f_{0}^{t},f_{0}^{s} \right){{\textbf{u}}_{i}}{\textbf{u}}_{i}^{H}{\textit{\textbf{a}}}\left( f_{0}^{t},f_{0}^{s} \right). \\ 
\end{aligned}
\end{equation}
With (31), we have the optimal estimate $\hat{h}_{i}^{*}$  as 
\begin{equation}
\begin{aligned}
\hat{h}_{i}^{*}\
&=\frac{{{{\hat{\rho }}}_{i}}+{{c}_{N}}}{\hat{\rho }_{i}^{2}+{{{\hat{\rho }}}_{i}}}
%\centerdot \\
& \quad \left[ \frac{\sum\limits_{q=1}^{Q}{\frac{{{c}_{N}}{{{\hat{k}}}_{q}}}{{{{\hat{\rho }}}_{q}}}}}{{{\textit{\textbf{a}}}^{H}}\left( f_{0}^{t},f_{0}^{s} \right){\textit{\textbf{a}}}\left( f_{0}^{t},f_{0}^{s} \right)-\sum\limits_{q=1}^{Q}{{{{\hat{k}}}_{q}}}+\sum\limits_{q=1}^{Q}{\frac{{{c}_{N}}{{{\hat{k}}}_{q}}}{\hat{\rho }_{q}^{2}}}}-{{{\hat{\rho }}}_{i}} \right] 
\end{aligned}
\end{equation}
and
\begin{equation}
\left| \tilde{h}_{i}^{*}-\hat{h}_{i}^{*} \right|\xrightarrow{a.s.}0.
\end{equation}
Then the estimated optimal inverse CNCM of ${{\bar{\textit{\textbf{R}}}}^{-1}}$  in the asymptotic case can be expressed as
\begin{equation}
{{\hat{\textit{\textbf{R}}}}^{-1}}\left( {{{\hat{\textbf{h}}}}^{*}} \right)=\frac{1}{\sigma _{n}^{2}}\left( \sum\limits_{i=1}^{Q}{\hat{h}_{i}^{*}{{\textbf{u}}_{i}}{\textbf{u}}_{i}^{H}}+{{\textbf{I}}_{NK}} \right)
\end{equation}
and the corresponding estimated weight vector is given as
\begin{equation}
{{\hat{\textit{\textbf{w}}}}^{opt}}\left( {{{\hat{{\textbf{h}}}}}^{*}} \right)=\frac{{{{\hat{\textit{\textbf{R}}}}}^{-1}}\left( {{{\hat{{\textbf{h}}}}}^{*}} \right){\textit{\textbf{a}}}\left( f_{0}^{t},f_{0}^{s} \right)}{{{\textit{\textbf{a}}}^{H}}\left( f_{0}^{t},f_{0}^{s} \right){{{\hat{\textit{\textbf{R}}}}}^{-1}}\left( {{{\hat{{\textbf{h}}}}}^{*}} \right){\textit{\textbf{a}}}\left( f_{0}^{t},f_{0}^{s} \right)}
\end{equation}
In computing the inverse CNCM (34), it is assumed to know the noise power $\sigma _{n}^{2}$, which can be obtained as in \cite{r54,r55,r56,r57}.

It should be noticed that in FD-STAP, the STAP performance becomes poor as the clutter DOFs  $Q$ increases \cite{r3,r4,r5,r6}. However, for the proposed RMT-FD-STAP, the clutter-related eigenvalues are corrected as in (34), and then the RMT-FD-STAP has better performance than the FD-STAP. 

\subsection{Reduced-dimension STAP Using RMT}

This subsection develops RD-STAP using RMT, RMT-RD-STAP. Our aim is to obtain the consistent estimation of ${\textit{\textbf{R}}}_{rd}^{-1}$  from the reduced-dimensional data (12). Let us decompose ${{\textit{\textbf{R}}}_{rd}}$  as
\begin{equation}
\begin{aligned}
{{\textit{\textbf{R}}}_{rd}}
& =\sigma _{n}^{2}\sum\limits_{i=1}^{M}{{{\gamma }_{i}}{{\textbf{e}}_{i}}{\textbf{e}}_{i}^{H}} \\ 
& =\sigma _{n}^{2}\sum\limits_{i=1}^{{{Q}_{rd}}}{{{\gamma }_{i}}{{\textbf{e}}_{i}}{\textbf{e}}_{i}^{H}}+\sigma _{n}^{2}\sum\limits_{i={{Q}_{rd}}+1}^{M}{{{\gamma }_{i}}{{\textbf{e}}_{i}}{\textbf{e}}_{i}^{H}}  
\end{aligned}
\end{equation}
where ${{Q}_{rd}}$  denotes local clutter DOFs \cite{r58} and ${{Q}_{rd}}<Q$, ${{\textbf{e}}_{i}}$  is the  $i$-th eigenvector with the corresponding eigenvalue  $\sigma _{n}^{2}{{\gamma }_{i}}$ and  ${{\gamma }_{1}}\ge {{\gamma }_{2}}\ge \cdots {{\gamma }_{M}}$. The leading eigenvalues   ${{\gamma }_{i}}$ $\left( i=1,\cdots,{{Q}_{rd}} \right)$ are related to clutter, while the others are related to noise.

It is seen from (36) that the ${{\textit{\textbf{R}}}_{rd}}$  does not have the spiked structure as ${\textit{\textbf{R}}}$  because the noise covariance matrix is not an identity one \cite{r58}. Then the technique in the last sub-section cannot be directly applicable to the estimate of the inverse CNCM ${{\textit{\textbf{R}}}_{rd}^{-1}}$. Note that the optimal RD-STAP is to minimize the output clutter-plus-noise power (17). Then the essence of the RD-STAP is to find the estimate of ${{\textit{\textbf{R}}}_{rd}^{-1}}$  which can reduce the output clutter-plus-noise power (17). It is found that the power (17) can be reduced with the noise-related covariance matrix replaced by an identity matrix. Although there are no theoretical guarantees, a number of simulations in Section IV have confirmed the assertion. For RMT-RD-STAP, in addition to three assumptions A.1$\sim$A.3 replaced by the corresponding parameters, we make the assumption A.4 as follows.

{\bf{A.4}} The noise component in (12) is taken to be the complex white Gaussian noise with the noise power $\sigma _{n}^{2}$.

Define ${{\beta }_{i}}={{\gamma }_{i}}-1$. Then under A.4, the matrix ${{\textit{\textbf{R}}}_{rd}}$  can be approximated as
\begin{equation}
{{\textit{\textbf{R}}}_{rd}}\approx \sigma _{n}^{2}\left( \sum\limits_{i=1}^{{{Q}_{rd}}}{{{\beta }_{i}}{{\textbf{e}}_{i}}{\textbf{e}}_{i}^{H}}+{{\textbf{I}}_{M}} \right)
\end{equation}
With the eigen-decomposition of the matrix ${{\textit{\textbf{R}}}_{Lrd}}$  as
\begin{equation}
{{\textit{\textbf{R}}}_{Lrd}}=\sigma _{n}^{2}\sum\limits_{i=1}^{M}{{{\alpha }_{i}}{{\textbf{d}}_{i}}{\textbf{d}}_{i}^{H}}
\end{equation}
where ${{\textbf{d}}_{i}}$  is the  $i$-th eigenvector with the corresponding eigenvalue $\sigma _{n}^{2}{{\alpha }_{i}}$  and ${{\alpha }_{1}}\ge \cdots \ge {{\alpha }_{M}}$, an inverse CNCM is reconstructed as
\begin{equation}
\bar{\textit{\textbf{R}}}_{rd}^{-1}=\frac{1}{\sigma _{n}^{2}}\sum\limits_{i=1}^{M}{{{\varsigma }_{i}}{{\textbf{d}}_{i}}{\textbf{d}}_{i}^{H}}
\end{equation}
where ${1}/{\sigma _{n}^{2}}{{\varsigma }_{i}}$  are the eigenvalues of  $\bar{\textit{\textbf{R}}}_{rd}^{-1}$ and the parameters ${{\varsigma }_{i}}$  are to be reconstructed. By A.4, setting ${{\varsigma }_{{{Q}_{rd}}+1}}=\cdots ={{\varsigma }_{M}}=1$, and letting ${{g}_{i}}={{\varsigma }_{i}}-1$, (39) can be re-expressed as
\begin{equation}
\bar{\textit{\textbf{R}}}_{rd}^{-1}({\textbf{g}})=\frac{1}{\sigma _{n}^{2}}\left( \sum\limits_{i=1}^{{{Q}_{rd}}}{{{g}_{i}}{{\textbf{d}}_{i}}{\textbf{d}}_{i}^{H}}+{{\textbf{I}}_{M}} \right)
\end{equation}
in which the inverse matrix  $\bar{\textit{\textbf{R}}}_{rd}^{-1}$ is implicitly expressed as a function of the vector ${\textbf{g}}={{[{{g}_{1}}\cdots {{g}_{{{Q}_{rd}}}}]}^{T}}$. Correspondingly, the adaptive weight vector $\bar{\textit{\textbf{w}}}_{rd}^{opt}$ is given as
\begin{equation}
\bar{\textit{\textbf{w}}}_{rd}^{opt}\left( {\textbf{g}} \right)=\frac{\bar{\textit{\textbf{R}}}_{rd}^{-1}\left( {\textbf{g}} \right){{\textit{\textbf{a}}}_{rd}}\left( f_{0}^{t},f_{0}^{s} \right)}{{\textit{\textbf{a}}}_{rd}^{H}\left( f_{0}^{t},f_{0}^{s} \right)\bar{\textit{\textbf{R}}}_{rd}^{-1}\left( {\textbf{g}} \right){{\textit{\textbf{a}}}_{rd}}\left( f_{0}^{t},f_{0}^{s} \right)}
\end{equation}
Then the RD-STAP problem under the spiked covariance model can be defined as
\begin{equation}
{{\textbf{g}}^{*}}=\underset{\textbf{g}}{\mathop{argmin}}\,{{P}_{rd}}\left( \bar{\textit{\textbf{w}}}_{rd}^{opt}\left( {\textbf{g}} \right) \right)
\end{equation}
where ${{P}_{rd}}\left( \bar{\textit{\textbf{w}}}_{rd}^{opt}\left( {\textbf{g}} \right) \right)$ is given by 
%(43) (see the top line of the next page).\\
%\begin{figure*}[!t]	
%	% ensure that we have normalsize text	
%	\normalsize	
%	% Store the current equation number.	
%	\setcounter{mytempeqncnt}{42}	
%	% Set the equation number to one less than the one	
%	% desired for the first equation here.	
%	% The value here will have to changed if equations	
%	% are added or removed prior to the place these	
%	% equations are referenced in the main text.	
%	%	\setcounter{equation}{24}
\begin{equation}
\begin{aligned}
& {{P}_{rd}}\left( \bar{\textit{\textbf{w}}}_{rd}^{opt}\left( {\textbf{g}} \right) \right)=\frac{{\textit{\textbf{a}}}_{rd}^{H}\left( f_{0}^{t},f_{0}^{s} \right)\bar{\textit{\textbf{R}}}_{rd}^{-1}\left( {\textbf{g}} \right){{\textit{\textbf{R}}}_{rd}}\bar{\textit{\textbf{R}}}_{rd}^{-1}\left( {\textbf{g}} \right){{\textit{\textbf{a}}}_{rd}}\left( f_{0}^{t},f_{0}^{s} \right)}{{{\left[ {\textit{\textbf{a}}}_{rd}^{H}\left( f_{0}^{t},f_{0}^{s} \right)\bar{\textit{\textbf{R}}}_{rd}^{-1}\left( {\textbf{g}} \right){{\textit{\textbf{a}}}_{rd}}\left( f_{0}^{t},f_{0}^{s} \right) \right]}^{2}}} \\ 
& \text{          }=\sigma _{n}^{2}\frac{{\textit{\textbf{a}}}_{rd}^{H}\left( f_{0}^{t},f_{0}^{s} \right)\left( {{\textbf{I}}_{M}}+\sum\limits_{q=1}^{{{Q}_{rd}}}{{{g}_{q}}{{\textbf{d}}_{q}}{\textbf{d}}_{q}^{H}} \right)\left( {{\textbf{I}}_{M}}+\sum\limits_{j=1}^{{{Q}_{rd}}}{{{\beta }_{j}}{{\textbf{e}}_{j}}{\textbf{e}}_{j}^{H}} \right)\left( {{\textbf{I}}_{M}}+\sum\limits_{i=1}^{{{Q}_{rd}}}{{{g}_{i}}{{\textbf{d}}_{i}}{\textbf{d}}_{i}^{H}} \right){{\textit{\textbf{a}}}_{rd}}\left( f_{0}^{t},f_{0}^{s} \right)}{{{\left[ {\textit{\textbf{a}}}_{rd}^{H}\left( f_{0}^{t},f_{0}^{s} \right)\left( {{\textbf{I}}_{M}}+\sum\limits_{i=1}^{{{Q}_{rd}}}{{{g}_{i}}{{\textbf{d}}_{i}}{\textbf{d}}_{i}^{H}} \right){{\textit{\textbf{a}}}_{rd}}\left( f_{0}^{t},f_{0}^{s} \right) \right]}^{2}}}  
\end{aligned}
\end{equation}
%\setcounter{equation}{43}
%% IEEE uses as a separator
%\hrulefill
%% The spacer can be tweaked to stop underfull vboxes.
%\vspace*{4pt}
%\end{figure*}
\quad Similar to the RMT-FD-STAP, we can obtain the optimal estimate $\hat{g}_{i}^{*}$ to minimize (43) as
\begin{equation}
\begin{aligned}
&\hat{g}_{i}^{*}=\frac{{{{\hat{\beta }}}_{i}}+{{c}_{Nrd}}}{\hat{\beta }_{i}^{2}+{{{\hat{\beta }}}_{i}}}
% \centerdot\\
& \left[ \frac{\sum\limits_{q=1}^{{{Q}_{rd}}}{\frac{{{c}_{Nrd}}{{{\hat{k}}}_{rdq}}}{{{{\hat{\beta }}}_{q}}}}}{a_{rd}^{H}\left( f_{0}^{t},f_{0}^{s} \right){{a}_{rd}}\left( f_{0}^{t},f_{0}^{s} \right)-\sum\limits_{q=1}^{{{Q}_{rd}}}{{{{\hat{k}}}_{rdq}}}+\sum\limits_{q=1}^{{{Q}_{rd}}}{\frac{{{c}_{Nrd}}{{{\hat{k}}}_{rdq}}}{\hat{\beta }_{q}^{2}}}}-{{{\hat{\beta }}}_{i}} \right]
\end{aligned}
\end{equation}
where ${{\hat{\beta }}_{i}}={\left[ {{\alpha }_{i}}-1-{{c}_{Nrd}}+\sqrt{{{\left( {{\alpha }_{i}}-1-{{c}_{Nrd}} \right)}^{2}}-4{{c}_{Nrd}}} \right]}/{2}$  is the consistent estimation of ${{\beta }_{i}}$, ${{c}_{Nrd}}={M}/{L}$, ${{\hat{k}}_{rdi}}=\frac{\left( 1+{{{c}_{Nrd}}}/{{{{\hat{\beta }}}_{i}}} \right)}{\left( 1-{{{c}_{Nrd}}}/{\hat{\beta }_{i}^{2}} \right)}{\textit{\textbf{a}}}_{rd}^{H}\left( f_{0}^{t},f_{0}^{s} \right){{\textbf{d}}_{i}}{\textbf{d}}_{i}^{H}{{\textit{\textbf{a}}}_{rd}}\left( f_{0}^{t},f_{0}^{s} \right)$.

Then the estimated optimal inverse CNCM can be expressed as
\begin{equation}
\hat{\textit{\textbf{R}}}_{rd}^{-1}\left( {{{\hat{\textbf{g}}}}^{*}} \right)=\frac{1}{\sigma _{n}^{2}}\left( \sum\limits_{i=1}^{{{Q}_{rd}}}{\hat{g}_{i}^{*}{{\textbf{d}}_{i}}{\textbf{d}}_{i}^{H}}+{{\textbf{I}}_{M}} \right)
\end{equation}
and the corresponding estimated optimal adaptive weight vector can be expressed as
\begin{equation}
\hat{\textit{\textbf{w}}}_{rd}^{opt}\left( {{{\hat{\textbf{g}}}}^{*}} \right)=\frac{\bar{\textit{\textbf{R}}}_{rd}^{-1}\left( {{{\hat{\textbf{g}}}}^{*}} \right){{\textit{\textbf{a}}}_{rd}}\left( f_{0}^{t},f_{0}^{s} \right)}{{\textit{\textbf{a}}}_{rd}^{H}\left( f_{0}^{t},f_{0}^{s} \right)\bar{\textit{\textbf{R}}}_{rd}^{-1}\left( {{{\hat{\textbf{g}}}}^{*}} \right){{\textit{\textbf{a}}}_{rd}}\left( f_{0}^{t},f_{0}^{s} \right)}
\end{equation}
Similar to the development of RMT-FD-STAP, the number of training samples must be not less than the local clutter DOFs, i.e. $L\ge {{Q}_{rd}}$.

As in the development of RMT-FD-STAP, the RMT-RD-STAP corrects the clutter-related eigenvalues by (44) and thus performs better than the RD-STAP for any local clutter DOFs ${{Q}_{rd}}$.

\subsection{Computational Cost}

As is well-known, there are two common ways in implementing STAP, the inversion or eigen-decomposition of the CNCM and the spatial-temporal filtering, which takes main computational costs \cite{r3}, \cite{r5}. Hence, we make a comparison of computational complexities of four different STAP algorithms, FD-STAP, RD-STAP, RMT-FD-STAP and RMT-RD-STAP. As discussed in last two sub-sections, both RMT-FD-STAP and RMT-RD-STAP are based on their respective sample CNCMs. Then RMT-RD-STAP/RMT-FD-STAP will have the same computational costs as RD-STAP/FD-STAP in terms of the common operations. Similar to RD-STAP, RMT-RD-STAP has less computational costs than FD-STAP. Of course, to compute the optimal weights by RMT-RD-STAP/RMT-FD-STAP, additional little computations are required as indicated in (34)/(45). Then in practical implementations, RMT-RD-STAP and RMT-FD-STAP will increase little computational costs.

\section{Numerical Results}

In this section, we conduct simulations to demonstrate the reasonability of the assumption A.4 and to evaluate the performances of the proposed RMT-FD-STAP and RMT-RD-STAP algorithms. The parameters used in the simulation scenarios are listed in Table 1. For comparisons, EFA with three Doppler channels is taken as an example of RD-STAP algorithms. In simulations, the noise power is $\sigma _{n}^{2}=1$, and the platform velocity  $V$ or the slop parameter  $\xi$ is set to be different values to reveal the performance of the proposed algorithms under different clutter DOFs. STAP is taken in the main lobe of antenna arrays.
\begin{table}[H]
\centering 
\caption{Parameters Used in the Simulations}
\begin{tabular}{ccc}
	\toprule
	Parameter &	Value &	Unit \\
	\midrule
	Height & 6000 & m \\
%	Range resolution & 100 & m \\
	Wavelength & 0.3 & m \\
	Array number & 8 & / \\
	Pulse number & 8 & / \\
	PRF & 2000 & Hz \\
	CNR & 30 & dB \\
	\bottomrule
\end{tabular}
\end{table}

\subsection{Simulation Analysis on the A.4}

\begin{figure}
	\centering%
	\subfigbottomskip=0pt
	\subfigcapskip=-5pt
	\subfigure[]{
		\includegraphics[width=17pc]{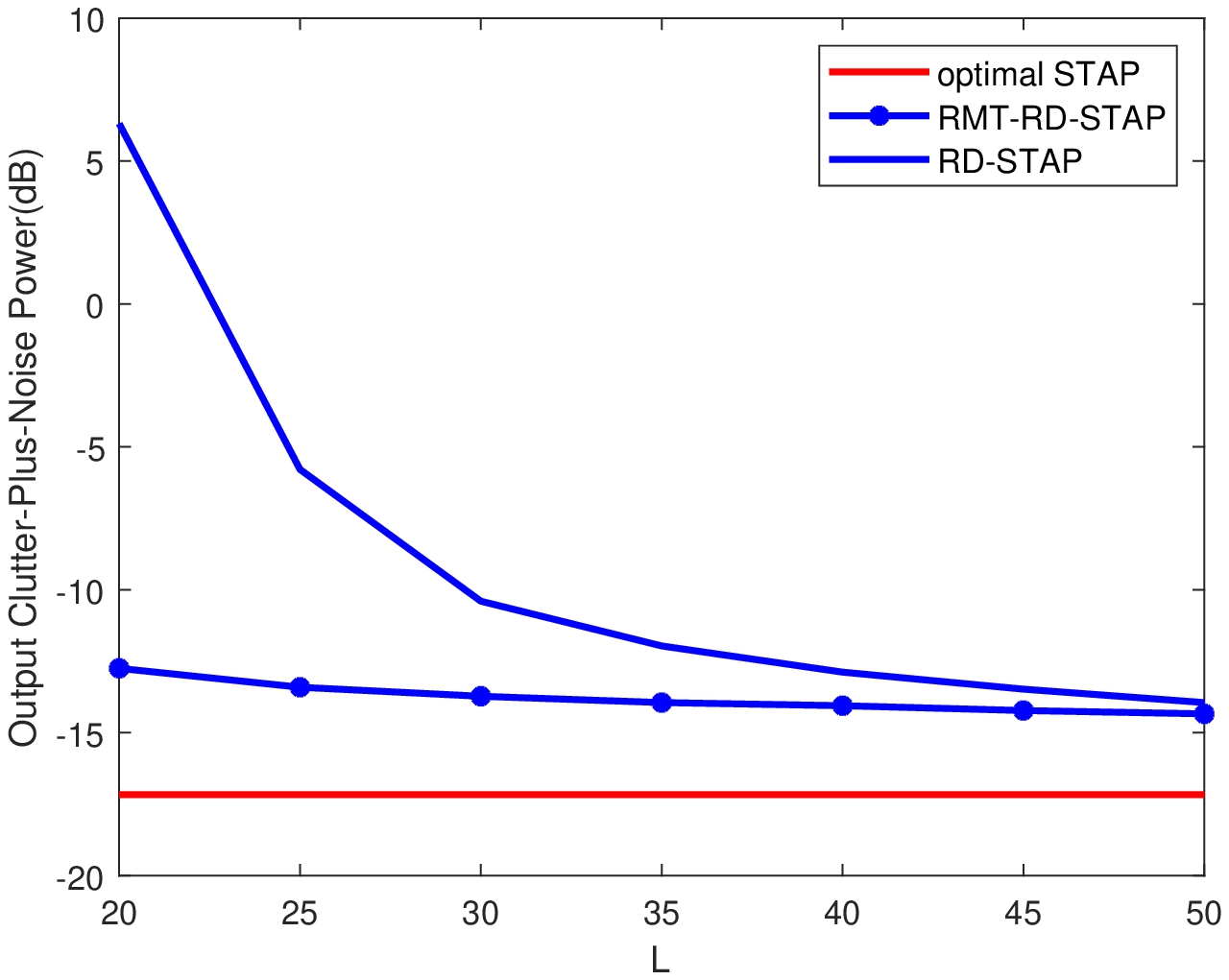}}
	\subfigure[]{
		\includegraphics[width=17pc]{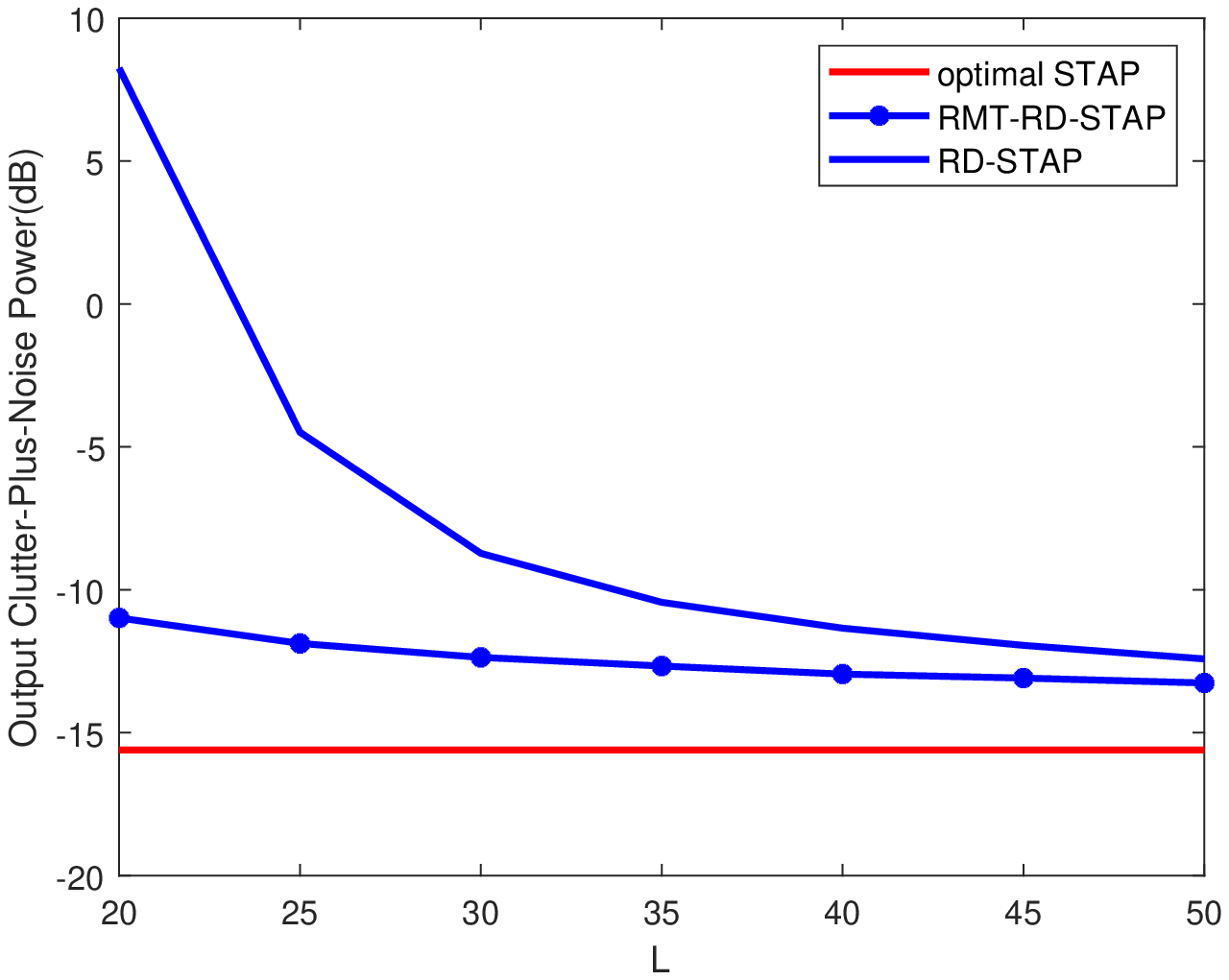}}\\
	\caption{The output clutter-plus-noise power against the number $L$ of the training samples.	(a) $V=150{m}/{s}$; (b) $V=300{m}/{s}$.	}
\end{figure}

We make a comparison of the output clutter-plus-noise powers derived by (7), (16) and (46) for two platform velocities, $V=150{m}/{s}\;$  and $V=300{m}/{s}$, corresponding to two local clutter DOFs, ${{Q}_{rd}}=10$  and ${{Q}_{rd}}=12$, respectively. Fig. 1 shows the output clutter-plus-noise power versus the number of training samples with $f_{0}^{t}=0.3$. It is obvious that for large $L$, both RD-STAP and RMT-RD-STAP have close output power. However, when the number $L$  of training samples becomes small, the output clutter-plus-noise power by the RD-STAP rapidly increases, while the power by the RMT-RD-STAP almost keeps intact. This simulation illustrates that the assumption of complex white Gaussian noise in reduced dimension is reasonable for the reduction of the output clutter-plus-noise power. By comparing Fig. 1(a) with Fig. 1(b), we can find that the local clutter DOFs have negative effects on the output clutter-plus-noise powers. The larger the local clutter DOFs are, the larger the optimal output powers are. 
But the gap between the output powers produced by RMT-RD-STAP and the optimal STAP changes slightly for these two clutter DOFs, which implies that RMT-RD-STAP is insensitive to clutter DOFs as demonstrated in the next subsection.

\subsection{Performance Comparisons}

Output SCNR loss, defined as the ratio of the output SCNR to the SNR achieved by a matched filter in a clutter-free environment \cite{r3}, is taken as a performance metric to measure different STAP algorithms. For FD-STAP, the SCNR loss is given as
\begin{equation}
SCNR_{Loss}^{FD-STAP}=\frac{\sigma _{n}^{2}}{NK}\frac{{{\left| {\textit{\textbf{w}}}{{_{L}^{opt}}^{\mathrm{H}}}{\textit{\textbf{a}}}\left( f_{0}^{t},f_{0}^{s} \right) \right|}^{2}}}{{\textit{\textbf{w}}}{{_{L}^{opt}}^{\mathrm{H}}}{\textit{\textbf{R}}}{\textit{\textbf{w}}}_{L}^{opt}}
\end{equation}	
Similar loss functions are given by optimal STAP, RD-STAP, RMT-FD-STAP and RMT-FD-STAP. In the following comparisons, the output SCNR losses are presented by averaging 1000 independent simulation runs.

Firstly, we present the variations of the output SCNR losses as the normalized Doppler frequencies for different numbers of training samples. The results are shown in Fig. 2. In the simulation, the platform is assumed to moves at the velocity $V=150{m}/{s}$. Then the slope parameter is $\xi =1$, and the clutter DOFs  $Q$ and  ${{Q}_{rd}}$ can be calculated respectively to be equal to 15 and 10. It is seen that STAP algorithms with RMT are superior to traditional STAP algorithms for small training samples. For Fig. 2(a), $L=10$, it is seen that RMT-RD-STAP is much superior to that of RD-STAP and FD-STAP and approaches mostly to the optimal performance. The SCNR loss by RMT-RD-STAP decreases about 20dB in comparison with RD-STAP. In this case, $L<Q$, RMT-FD-STAP is not applicable. For Fig. 2(b), $L=15$, the SCNR losses by simulated algorithms decrease in comparison with Fig. 2(a). RMT-RD-STAP is still superior to RD-STAP and reduces SCNR loss about 15dB. RMT-FD-STAP is superior to both FD-STAP and RD-STAP. Because of insufficient training samples, RMT-RD-STAP is superior to RMT-FD-STAP. For Fig. 2(c), $L=48$  and the number of training samples satisfies the RMB rule in reduced dimensionality. Three algorithms by RMT-FD-STAP, RMT-RD-STAP and RD-STAP have good performance and approach to the optimal one. Among them, RMT-RD-STAP is superior to RD-STAP, and RMT-FD-STAP has the best performance because of sufficient training samples. FD-STAP has the worst performance due to much less training samples. For Fig. 2(d), $L=128$. It is seen that all STAP algorithms perform better because of sufficient training samples. It is interesting to note that RD-STAP is superior to RMT-RD-STAP. This is because RD-STAP can achieve almost optimal performance for such training samples, while RMT-RD-STAP exploits approximate operations in deriving the FD-STAP problem (43). 
\begin{figure}
	\centering%
	\subfigbottomskip=0pt
	\subfigcapskip=-5pt
	\subfigure[]{
	\includegraphics[width=17pc]{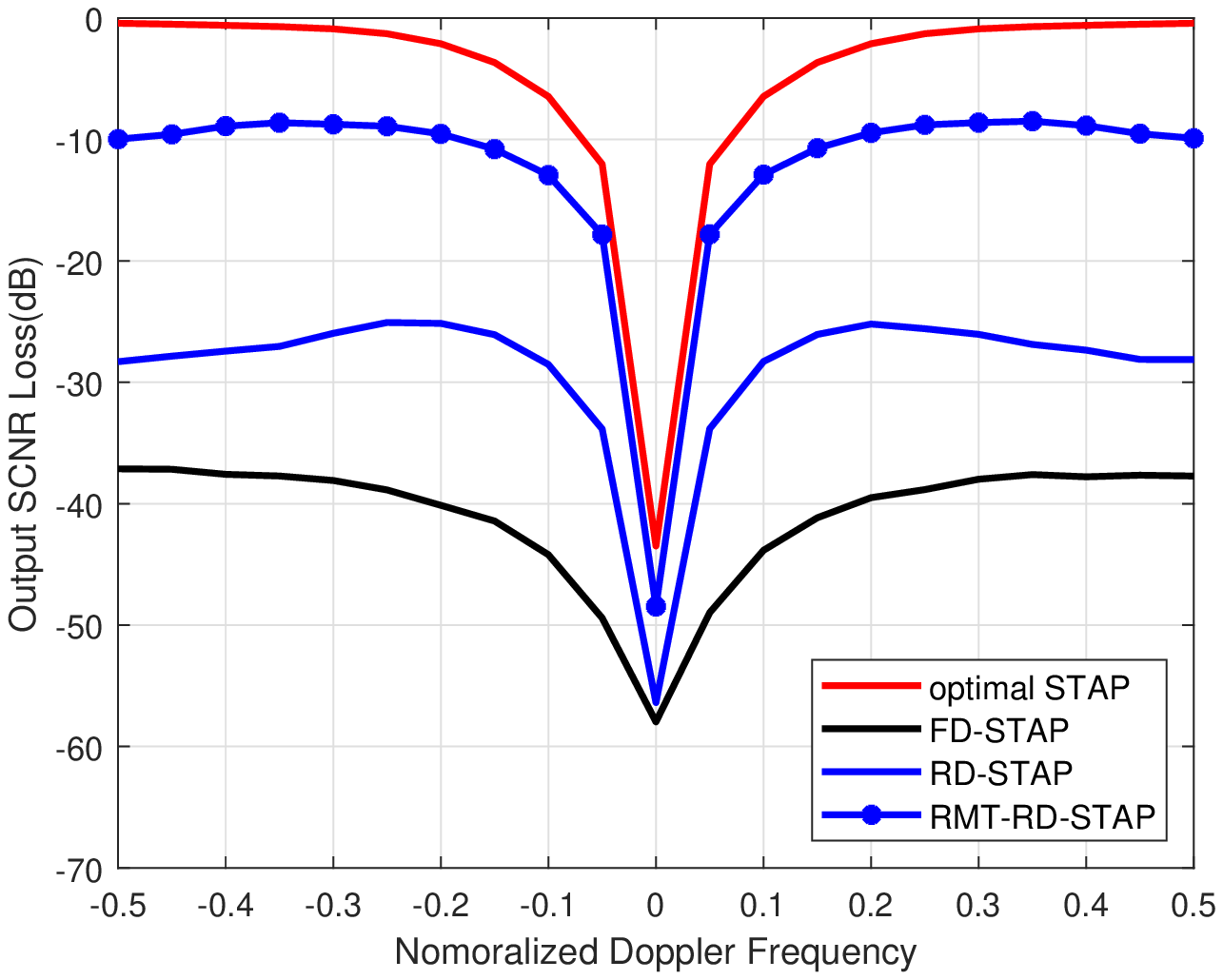}}
	\subfigure[]{
	\includegraphics[width=17pc]{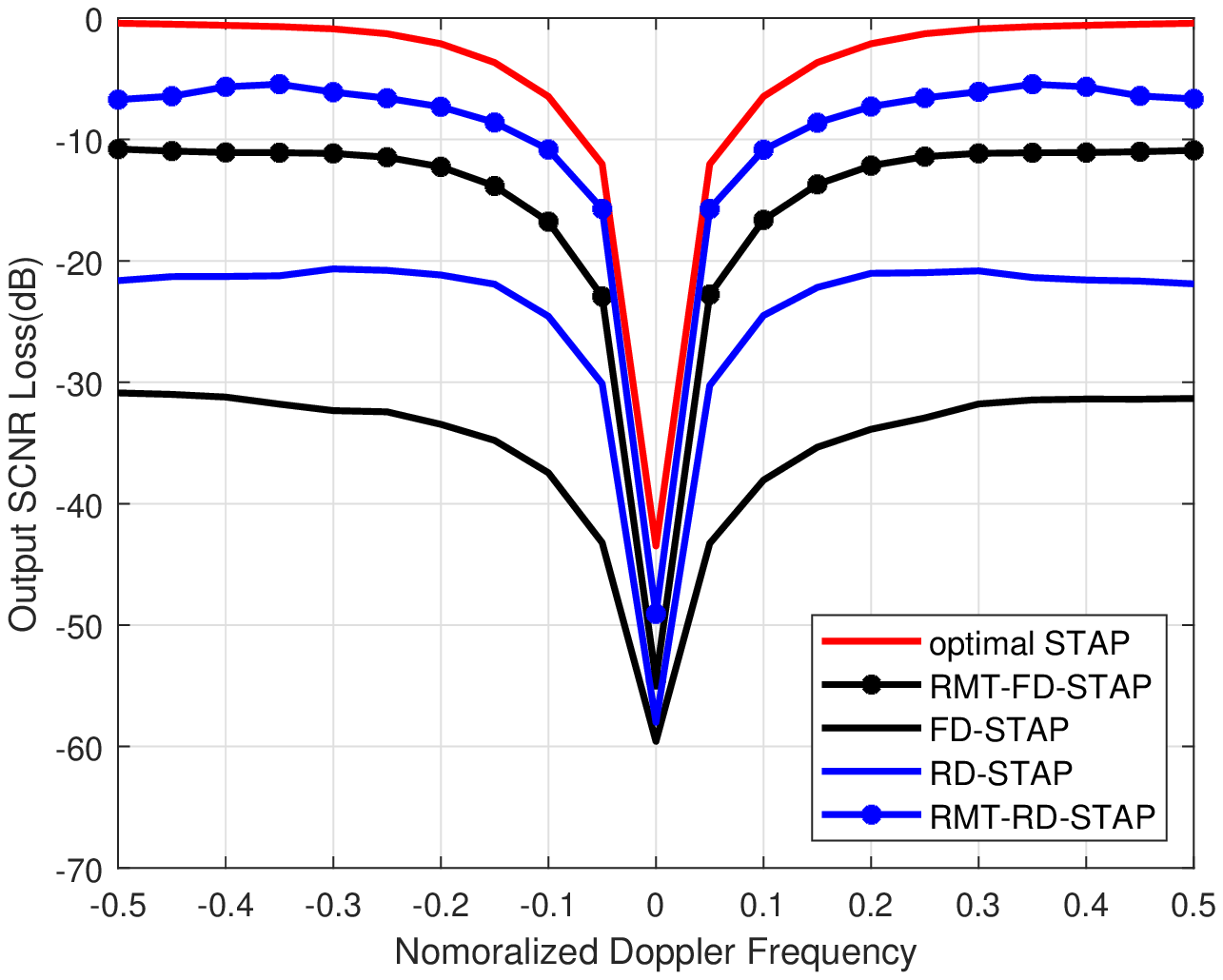}}\\
	\subfigure[]{
	\includegraphics[width=17pc]{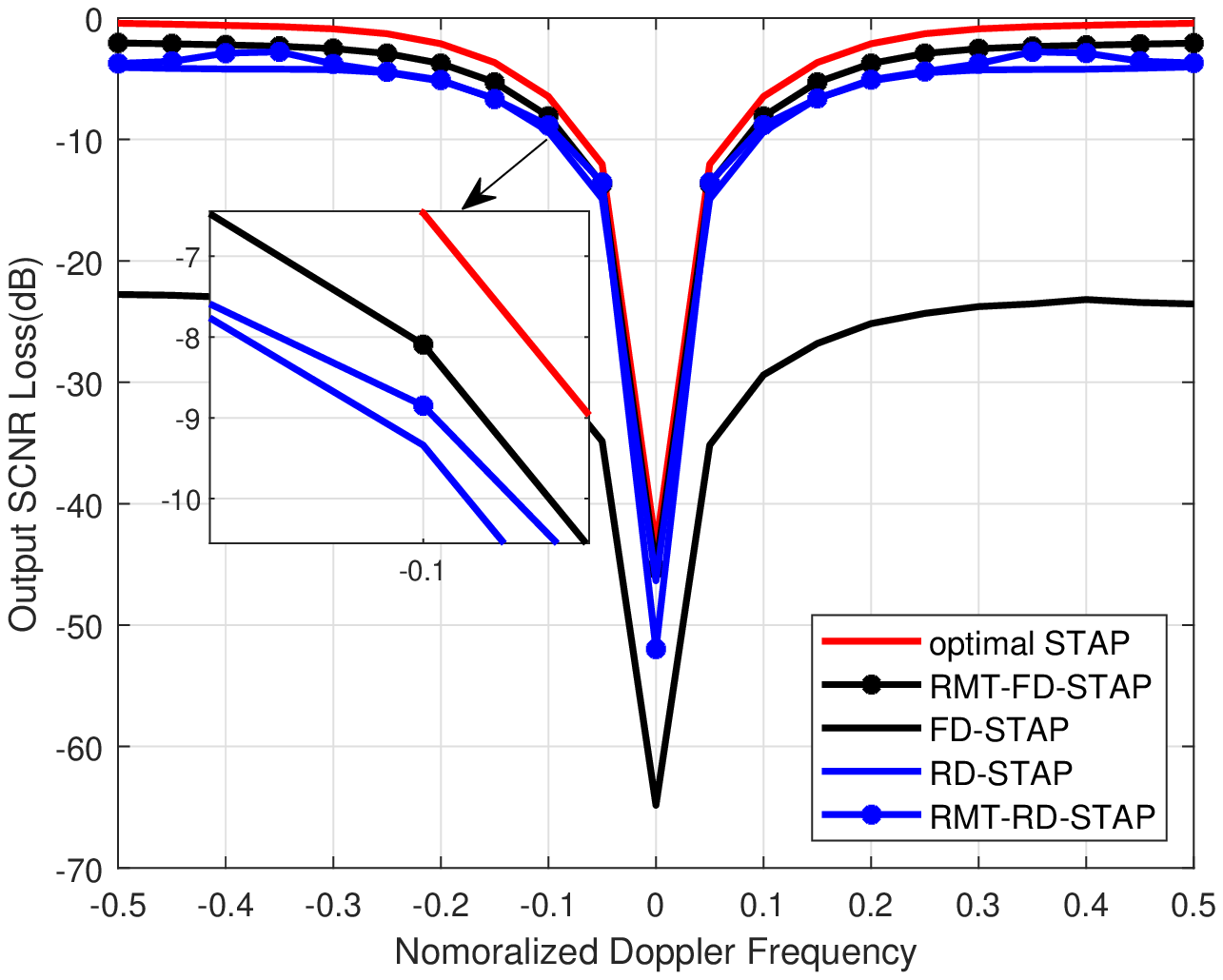}}
	\subfigure[]{
	\includegraphics[width=17pc]{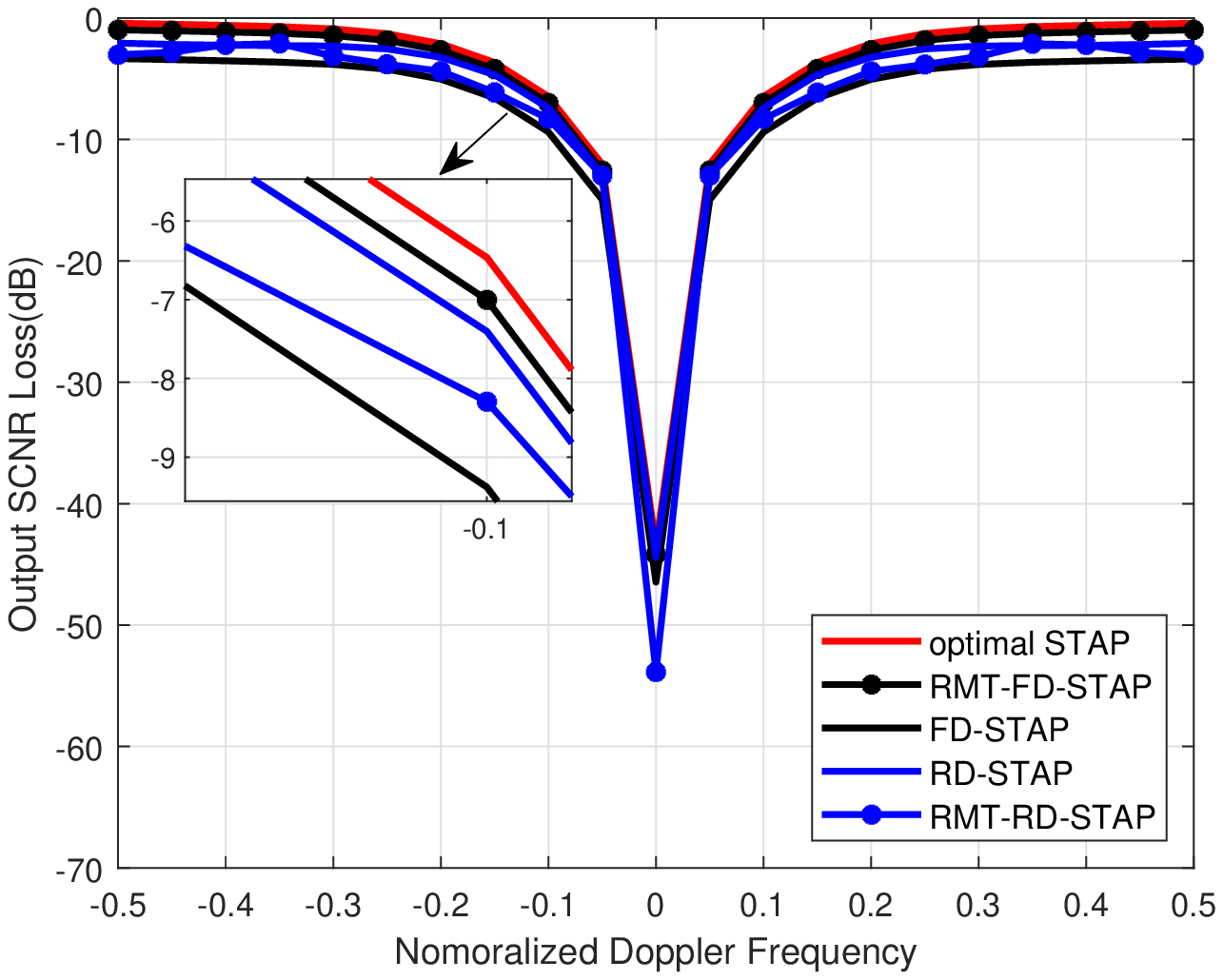}}\\
    \caption{Output SCNR losses versus normalized Doppler frequencies. (a) 10 training samples; (b) 15 training samples; (c) 48 training samples; (d) 128 training samples.}
\end{figure}

Figure 3 shows the output SCNR losses versus the number of the training samples with $f_{0}^{t}=0.3$. In the simulated range of the training samples, it is seen that for large $L$, RD-STAP, RMT-FD-STAP and RMT-RD-STAP achieve the performance close to each other and are much superior to that of FD-STAP. However, as the number of the training samples decreases, RMT-FD-STAP and RMT-RD-STAP have distinct performance advantages over FD-STAP and RD-STAP. Anyway, FD-STAP performs badly. Along with Fig. 2, Fig. 3 further demonstrates the advantages of the proposed STAP algorithms.

\begin{figure}
	\centering{\includegraphics[width=17pc]{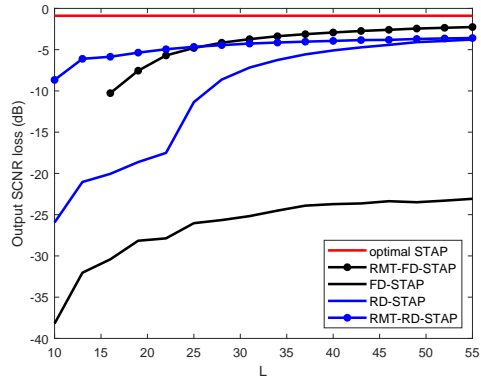}}
	\caption{Output SCNR losses versus the number $L$ of the training samples.}
\end{figure}
\begin{figure}
	\centering%
	\subfigbottomskip=0pt
	\subfigcapskip=-5pt
	\subfigure[]{
		\includegraphics[width=17pc]{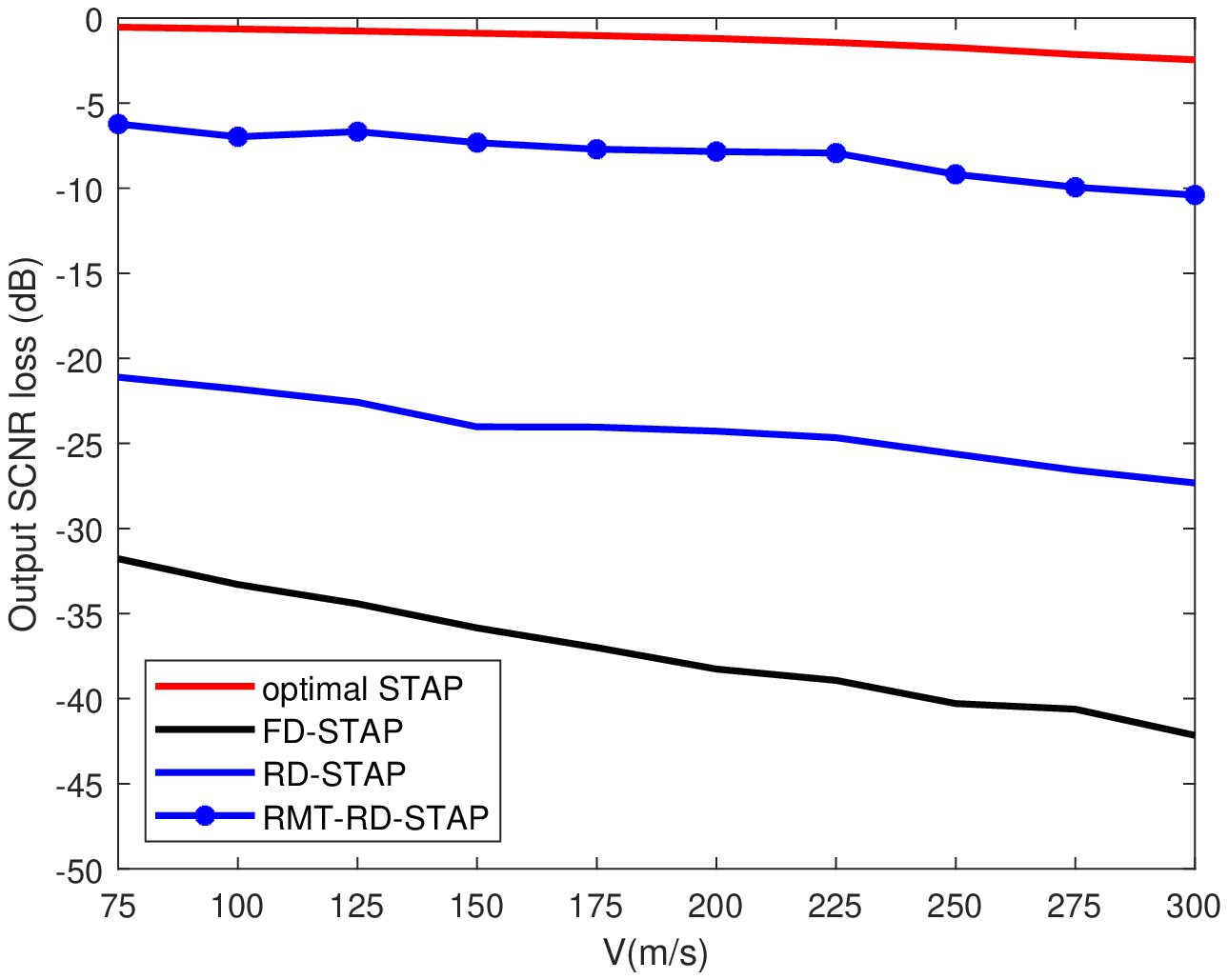}}
	\subfigure[]{
		\includegraphics[width=17pc]{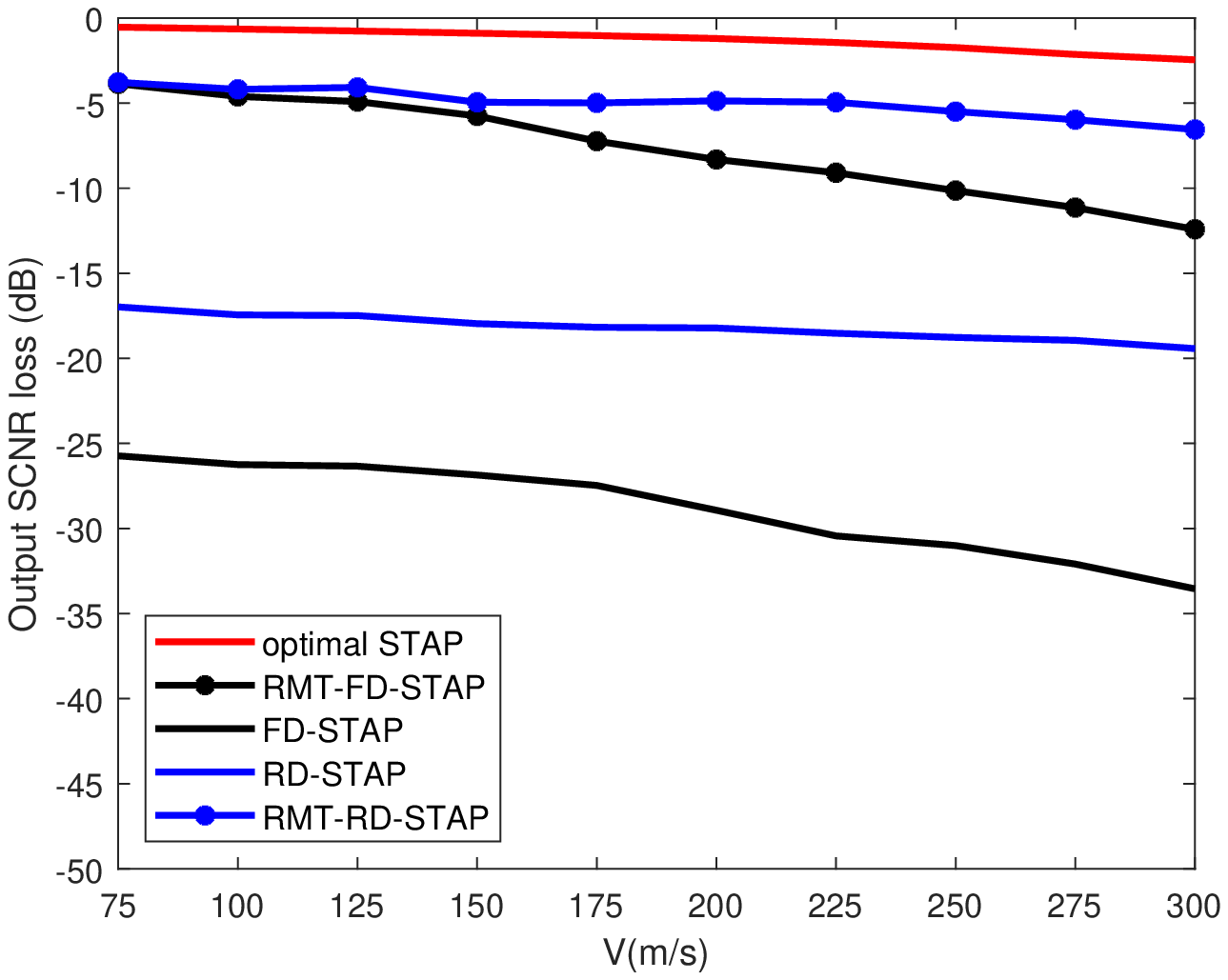}}\\
	\caption{Output SCNR losses versus velocities $V$. (a) 12 training samples; (b) 22 training samples. }
\end{figure}

Next, we present the output SCNR losses versus the platform velocities to demonstrate the performance advantages of the proposed algorithms under different clutter DOFs. Fig. 4 shows the simulation results for $f_{0}^{t}=0.3$. The clutter DOFs  $Q$ and  ${{Q}_{rd}}$ can be calculated respectively as $Q=22$  and ${{Q}_{rd}}=12$  for the maximum platform velocity  $V=300{m}/{s}$. Then we set two training samples $L=12$ and $L=22$. In such cases, both FD-STAP and RD-STAP do not satisfy their respective RMB rules. It is seen that RMT-FD-STAP and RMT-RD-STAP have great performance advantages in comparison with FD-STAP and RD-STAP for the small training samples. For Fig. 4(a) with $L=12$,  $L$ is less than the clutter DOFs in full dimension for the platform velocity larger than  $60{m}/{s}$, and RMT-FD-STAP is not applicable. However, for the setting $L=12$, RMT-RD-STAP can work in the simulated velocity range. Then the performance improvement in Fig. 2(a) can be achieved for all platform velocities. Similarly, the improvement in Fig. 2(b) can be achieved as indicated in Fig. 4(b).

In the implementation of the proposed STAP algorithms, the clutter DOFs need to be calculated according to radar parameters and platform parameters \cite{r3}, \cite{r58}. In real environment, these parameters often deviate from the designed ones and then the calculated clutter DOFs are different from the real clutter DOFs. To demonstrate the robustness of the proposed algorithms, we use Fig. 5 to show the output SCNR losses versus the clutter DOF errors for  $f_{0}^{t}=0.3$. In the simulation, the platform velocity is assumed to be  $V=150{m}/{s}$, and then the calculated clutter DOFs in full dimension and reduced dimension are 15 and 10, respectively. The clutter DOF errors vary from -3 to 3. The numbers of training samples are set to be  $L=18$ and  $L=13$ for the maximum clutter DOF error, respectively. In such cases, both FD-STAP and RD-STAP do not satisfy their respective RMB rules. It is observed from Fig. 5 that the proposed STAP algorithms, RMT-FD-STAP and RMT-RD-STAP, are not sensitive to the DOF errors, and have stable output SCNR losses, performing like traditional FD-STAP and RD-STAP algorithms. 
\begin{figure}
	\centering%
	\subfigbottomskip=0pt
	\subfigcapskip=-5pt
	\subfigure[]{
		\includegraphics[width=17pc]{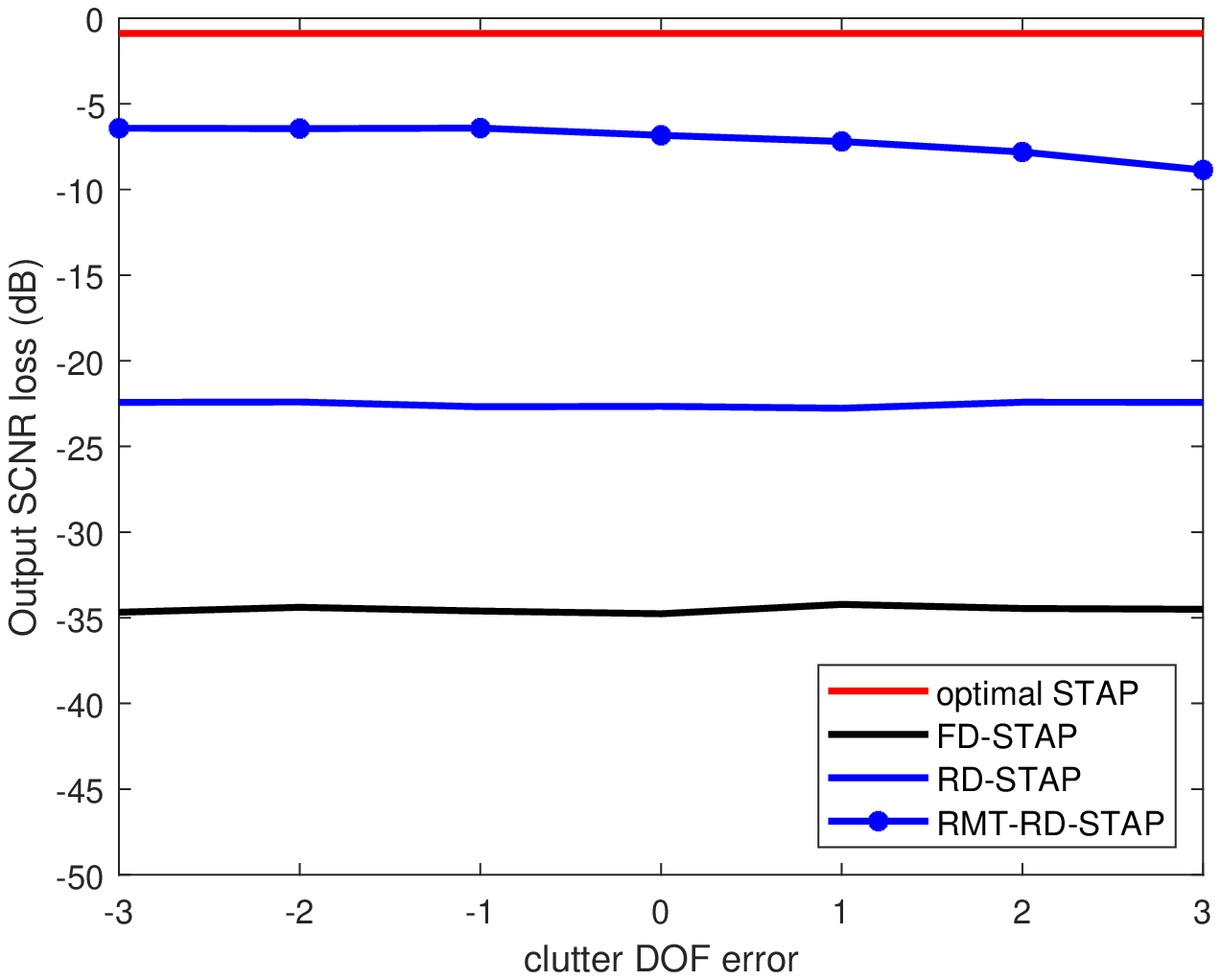}}
	\subfigure[]{
		\includegraphics[width=17pc]{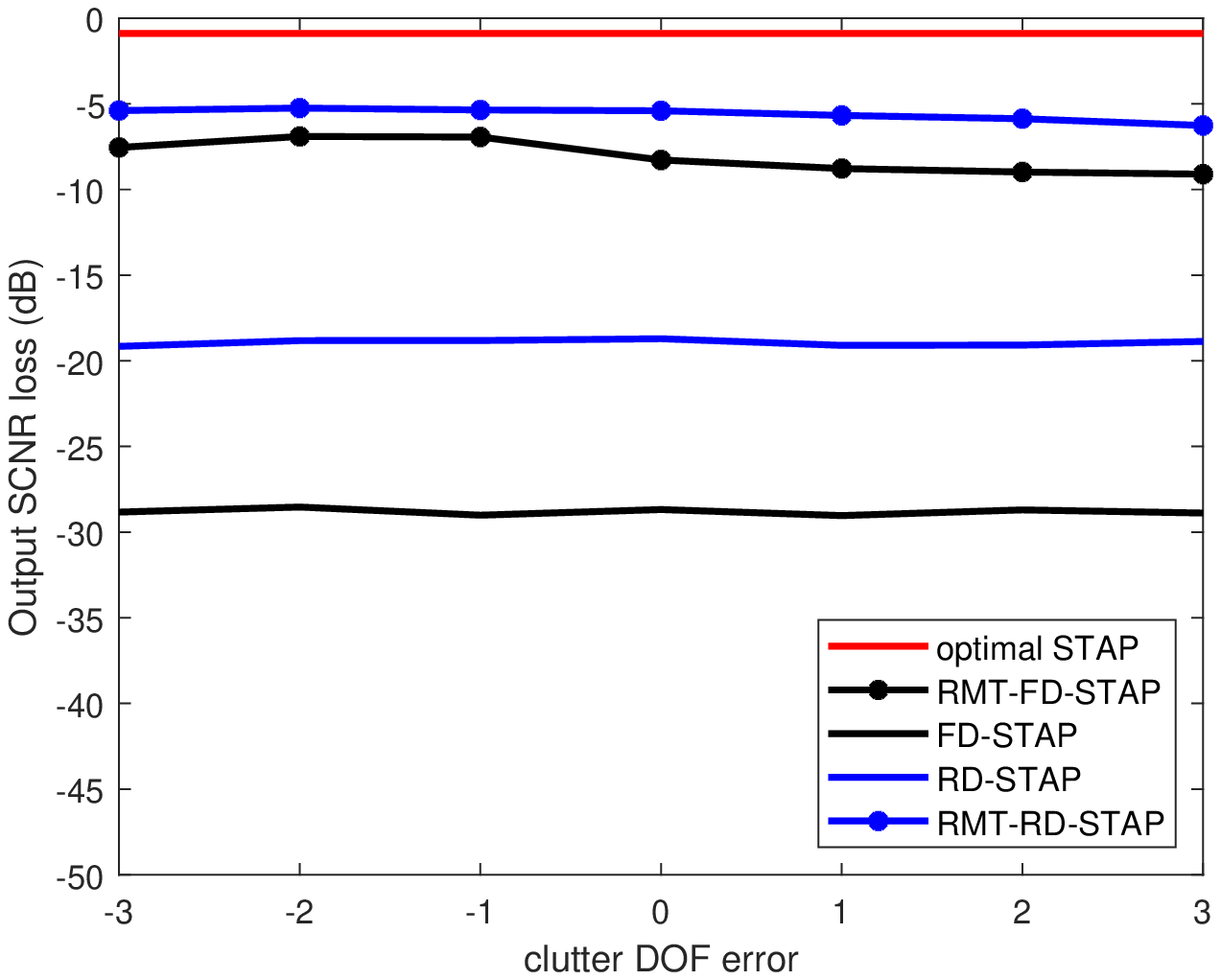}}\\
	\caption{Output SCNR losses versus clutter DOF errors. (a) 13 training samples; (b) 18 training samples.	}
\end{figure}
\section{Conclusions}
In this paper, we introduce a STAP theory with RMT under limited training samples. Two new STAP algorithms, RMT-FD-STAP and RMT-RD-STAP, are developed. Instead of directly taking the inversion of CNCM constructed from the maximum likelihood principle in the STAP development, this new theory in the work estimates the inversion by optimally manipulating its eigenvalues. By using RMT and the spiked covariance models, the proposed STAP algorithms greatly outperform other related algorithms with small clutter training samples and have superior performance in the case of the training samples satisfying the RMB rule.

The future work will include the further performance analyses and verification of the proposed STAP algorithms.

\bibliographystyle{IEEEtran}
\bibliography{myref}

\end{document}